\documentclass[aps,final,onecolumn,superscriptaddress,nofootinbib,floatfix,11pt]{revtex4-1}
\usepackage{amsfonts,amssymb,stmaryrd,latexsym,amsmath,braket,comment}
\usepackage{graphicx,subfigure}
\usepackage{times}
\usepackage{slashed}

\def\bm{\boldsymbol}

\def\ra{\rangle}

\def\vn{{\bm n}}
\def\vp{{\bm p}}

\newcommand{\bea}{\begin{eqnarray}}
\newcommand{\eea}{\end{eqnarray}}
\newcommand{\no}{\nonumber \\}

\begin{document}

\title{Quantum Many-Body Calculations using Body-Centered Cubic Lattices}

\author{Young-Ho Song}
\author{Youngman Kim}
\affiliation{Rare Isotope Science Project, Institute for Basic Science, Daejeon 34000, Korea}

\author{Ning~Li}
\affiliation{School of Physics, Sun Yat-Sen University, Guangzhou 510275, China}

\author{Bing-Nan~Lu}
\affiliation{China Academy of Engineering Physics, Graduate School, Building 8, No.~10 Xi'er Road, ZPark II, Haidian District, Beijing, 100193, China}

\author{Rongzheng~He}
\author{Dean~Lee}
\affiliation{Facility for Rare Isotope Beams and Department of Physics and
Astronomy,
  Michigan State University, East Lansing, MI~48824, USA}

\begin{abstract}
It is often computationally advantageous to model space as a discrete set of points forming a lattice grid.
This technique is particularly useful for computationally difficult problems such as quantum many-body systems.  For reasons of simplicity and familiarity, nearly all quantum many-body calculations have been performed on simple cubic lattices.  Since the removal of lattice artifacts is often an important concern, it would be useful to perform calculations using more than one lattice geometry.  In this work we show how to perform quantum many-body calculations using auxiliary-field Monte Carlo simulations on a three-dimensional body-centered cubic (BCC) lattice.  As a benchmark test we compute the ground state energy of 33 spin-up and 33 spin-down fermions in the unitary limit, which is an idealized limit where the interaction range is zero and scattering length is infinite.  As a fraction of the free Fermi gas energy $E_{\rm FG}$, we find that the ground state energy is $E_0/E_{\rm FG}= 0.369(2), 0.371(2),$ using two different definitions of the finite-system energy ratio.  This is in excellent agreement with recent results obtained on a cubic lattice \cite{He:2019ipt}.  We find that the computational effort and performance on a BCC lattice is approximately the same as that for a cubic lattice with the same number of lattice points.  We discuss how the lattice simulations with different geometries can be used to constrain the size lattice artifacts in simulations of continuum quantum many-body systems.
\end{abstract}

\maketitle

\renewcommand{\baselinestretch}{1.0}

\section{Motivation}

While real physical systems reside in continuous space, it is often computationally advantageous to model space as a discrete set of points forming a lattice grid.  This is especially useful for computationally heavy calculations such as the case for quantum systems with many particles.  Lattice simulations have been used to study a wide range of different phenomena in quantum many-body systems. For example, lattice simulations in low-energy nuclear physics have been used to describe nuclear forces, structure, reactions, and thermodynamics \cite{Lee:2004si,Borasoy:2006qn,Abe:2007fe,Lee:2008fa,Drut:2012md,Wlazlowski:2014jna,Elhatisari:2015iga,Elhatisari:2016owd,Korber:2017emn,Elhatisari:2017eno,Lu:2018bat,Lahde:2019npb,Lu:2019nbg,Alexandru:2020zti}. In such calculations, a question of paramount concern is whether the systematic errors induced by the lattice grid can be properly removed from the final results.  

\begin{figure}[h]
                \centering               \includegraphics[width=5cm]{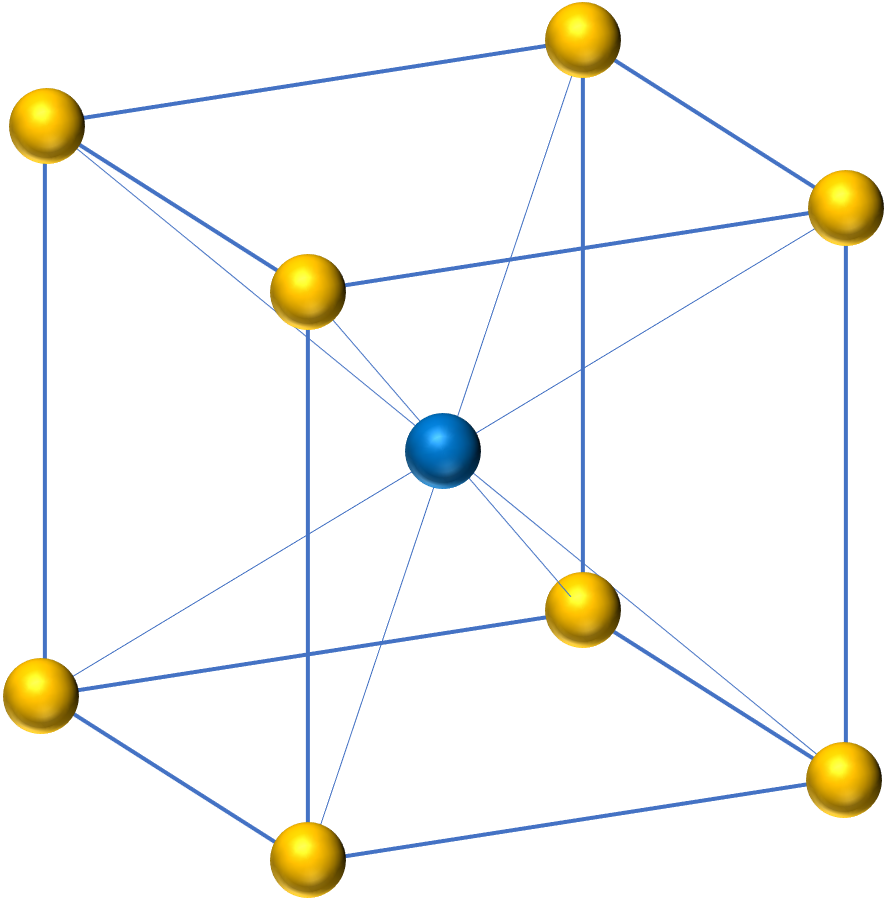}\\
                \caption{Drawing of one unit cell of the BCC lattice.}
                \label{BCC_lattice}
\end{figure}

In order to verify independence upon the lattice geometry, it would clearly be useful to perform calculations using more than one lattice geometry.  For reasons of simplicity and familiarity, however, nearly all quantum many-body calculations have been performed on simple cubic lattices.  In this paper we broaden the menu of computational lattice options and show how to perform quantum many-body calculations on a three-dimensional body-centered cubic (BCC) lattice.  The BCC lattice is convenient because it preserves the octahedral symmetry of the simple cubic lattice.  Having this large rotational symmetry group is helpful when approximating the full rotational symmetry of continuous space.  There are, however, some differences with the simple cubic lattice. Instead of one lattice site per unit cell, it has two lattice sites per unit cell.  And instead of six nearest neighbor lattice sites, it has eight nearest neighbors.  We show one unit cell of the BCC lattice in Fig.~\ref{BCC_lattice}.

In the literature one can find several studies of various spin models on BCC lattices \cite{Guttman:1961,Adler:1993kp,Caselle:1993jh,Butera:1995kr,Butera:1997ak,Butera:1998rk,Butera:1999qa,Campostrini:2000yf,Butera:2000zt,Butera:2018kjb,Radicevic:2019vyb,Shirley:2019,Murtazaev:2020,You:2020}.  To the best of our knowledge, BCC lattices have not yet been used for calculations of quantum many-body systems involving fermions.  The main goal of this paper is to show how simulations of quantum many-body systems with fermions can be realized on a BCC lattice.  We illustrate with a well-studied but nontrivial quantum many-body system called the unitary Fermi gas.

\section{Unitary Fermi gas}
The unitary Fermi gas refers to an interacting system of two-component fermions with zero range interactions and infinite scattering length.  The ground state of the unitary Fermi gas is a superfluid that sits in the crossover region between a weakly-coupled Bardeen-Cooper-Schrieffer (BCS) superfluid and strongly-coupled Bose-Einstein condensate (BEC).  Much of the nuclear physics interest comes from the fact that the unitary Fermi gas approximately describes the physics of dilute neutron matter found in the inner crust of a neutron star.  It is in that region that the interparticle spacing between neutrons is larger than the range of the neutron-neutron interactions (about 3 fm) but smaller than the neutron-neutron scattering length (about 19 fm).

The unitary Fermi gas is a scale-invariant system with no intrinsic length scale.  We can therefore use simple dimensional analysis to determine the scaling of observables with respect to the Fermi momentum $k_F$.  In particular, the ground state energy of the unitary Fermi gas must have the form
\begin{equation}
    E_0 = \xi E_{\rm FG},
\end{equation}
where $\xi$ is a universal dimensionless constant and $E_{\rm FG}$ is the ground state energy of the non-interacting Fermi gas.  The constant $\xi$ is sometimes called the Bertsch parameter and has been measured in many experiments using ultracold trapped atoms \cite{O'Hara:2002,Partridge:2006,Gehm:2003,Bourdel:2004,Bartenstein:2004,Kinast:2005,Regal:2005,Stewart:2006,Luo:2009,Navon:2010,Nascimbene:2010,Ku:2012}.  It has also been calculated by analytical methods \cite{Engelbrecht:1997,Baker:1999,Steele:2000,Heiselberg:2001,Strinati:2004,Schafer:2005,Papenbrock:2005,Nishida:2006,Haussman:2007,Veillette:2007,Arnold:2007,Chen:2007,Nishida:2009}. In addition, a substantial number of numerical studies have been made using lattice and continuum quantum Monte Carlo simulations as well as other techniques \cite{Endres:2012cw,Jensen:2019zkr,Mihaila:2011pq,Carlson:2003,Chang:2004,Astrakharchik:2004,Carlson:2005,Lee:2006a,Lee:2006b,Bulgac:2006,Lee:2006c,Juillet:2007,Lee:2008a,Lee:2008b,Li:2011,Bulgac:2008,Abe:2009,Magierski:2008wa,Gandolfi:2011,Endres:2012cw,Forbes:2010gt,Carlson:2011kv}.  

In this work, we compute $\xi$ for $33$ spin-up and $33$ spin-down fermions using a BCC lattice.  We will compare the performance and results with recent lattice calculations for the same system of $33$ spin-up and $33$ spin-down fermions using a simple cubic lattice~\cite{He:2019ipt}.  To facilitate a direct comparsion, we follow as closely as possible the same lattice formalism and methods used in Ref.~\cite{He:2019ipt}.


\section{Lattice action}
As we can see in Fig.~\ref{BCC_lattice}, there are two lattice sites per unit cell of the BCC lattice, and the length of the unit cell is what we call our lattice spacing $a_{\rm latt}$.  In our lattice calculations, we use cubic periodic boundary conditions with $L \times L \times L$ unit cells.  Unless otherwise indicated, we use units where factors of $a_{\rm latt}$ and $\hbar$ are not explicitly written.  The BCC lattice can be viewed as two simple cubic sublattices, both with lattice spacing $a_{\rm latt}$, such that the sites of one sublattice are positioned at the centers of the lattice cubes of the other sublattice and vice versa.  The points of the first sublattice are located at integer triples $(n_x,n_y,n_z)$, while the points of the second sublattice are shifting by $\left(\frac{1}{2},\frac{1}{2},\frac{1}{2}\right)$ from the first sublattice points.  The total number of lattice points is $2L^3$.

We denote the full set of lattice sites as ${\bm n}$, while the first sublattice is written as ${\bm n}_1$ and the second sublattice is written as ${\bm n}_2$.  The mass of all our fermions is $m$ and our lattice time step, $a_t$, is taken to be $0.0107\,ma_{\rm latt}^2\hbar^{-1}$.
The lattice Hamiltonian consists of free and interaction parts,
\bea
H= H_{\rm free} +V.
\eea
The free Hamiltonian takes the form
\bea
H_{\rm free}&=&\frac{\hbar^2}{2m}\sum_{\vn}
\left[ w_0a^\dagger(\vn)a(\vn)
+w_1\sum_{{\bm \Delta \in S_{\bm \Delta}}}a^\dagger(\vn+{\bm \Delta})a(\vn)
\right. \no & & \left. \quad 
+w_2\sum_{{\bm \Delta \in S_{\bm \Delta}}}a^\dagger(\vn+2{\bm \Delta})a(\vn)
+w_3\sum_{{\bm \Delta \in S_{\bm \Delta}}}a^\dagger(\vn+3{\bm \Delta})a(\vn)+\cdots
\right]\, , 
\eea
where the eight displacement vectors in $S_{\bm \Delta}$ are
\bea
S_{\bm \Delta}= \left(\pm \frac{1}{2},\pm \frac{1}{2},\pm \frac{1}{2}\right).
\eea

The eigenstates of our free Hamiltonian are momentum eigenstates $|\vp\ra$ with
momenta $\vp=(p_x,p_y,p_z)$ such that
\bea
H_{\rm free}|\vp\ra = \frac{\hbar^2}{2m} f(\vp) |\vp\ra,
\eea
where
\bea
f(\vp)&=& w_0
+8w_1 
\cos\left(\frac{p_x}{2}\right)
\cos\left(\frac{p_y}{2}\right)
\cos\left(\frac{p_z}{2}\right)\no
& & 
+8w_2 
\cos\left( p_x \right)
\cos\left( p_y\right)
\cos\left( p_z \right)\no
& & 
+8w_3
\cos\left( \frac{3p_x}{2}\right) 
\cos\left( \frac{3p_y}{2}\right) 
\cos\left( \frac{3p_z}{2}\right) +\cdots\, .
\eea
The values of the coefficients $w_i$ are fixed such that $f(\vp)$ gives the desired $p^2$ behavior up to some prescribed order in powers of the momentum.  At ${\cal O}(p^2)$, we have 
\bea
w_0=8, \quad w_1=-1 ,\quad w_2 =0,\quad w_3=0.
\eea
At ${\cal O}(p^4)$,
\bea
w_0=10, \quad w_1=-\frac{4}{3} ,\quad w_2 =\frac{1}{12},\quad w_3=0.
\eea
At ${\cal O}(p^6)$,
\bea
w_0=\frac{98}{9}, \quad w_1=-\frac{3}{2} ,\quad w_2 =\frac{3}{20},\quad w_3=-\frac{1}{90}.
\eea
In this work, we use the free Hamiltonian valid up to ${\cal O}(p^6)$.

\begin{figure}
                \centering
              \includegraphics[width=12cm]{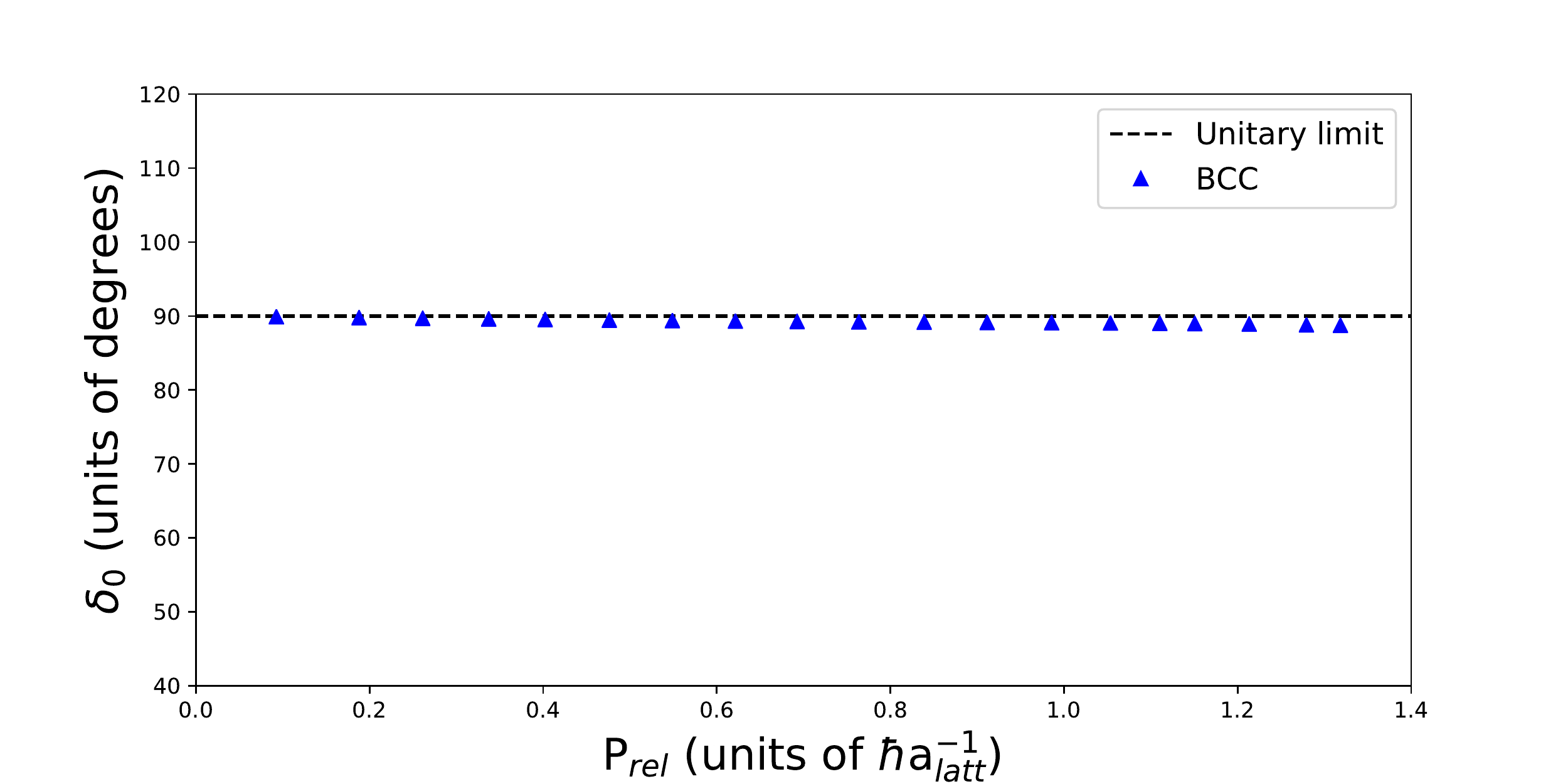}\\
                \caption{S-wave phase shifts computed on the BCC lattice versus relative momentum. The dashed line marks the unitary limit, which corresponds to a constant phase shift equal to $90$ degrees.}
                \label{phase_shift}
\end{figure}

For the lattice interaction we use the nonlocal smearing method described in Ref.~\cite{He:2019ipt}. It has the advantage of allowing the tuning of the S-wave scattering phase shifts as desired without introducing interactions in other partial wave channels.  The interaction has the form
\bea
V=\frac{C_0}{2}\sum_{\vn} : \rho_{\rm NL}(\vn)\rho_{\rm NL}(\vn): ,
\eea
where $::$ denotes normal ordering, where the annihilation operators are on the right and the creation operators are on the left.
The nonlocal density operator is defined by
\bea
\rho_{\rm NL}(\vn)&=&\sum_{\sigma=\uparrow,\downarrow} a^\dagger_{\sigma,{\rm NL}}(\vn)a_{\sigma,{\rm NL}}(\vn),
\eea
where
\bea
a_{\sigma,{\rm NL}}(\vn)&=& a_\sigma(\vn)+s_{\rm NL} \sum_{{\bm \Delta} \in S_{\bm \Delta}} a_\sigma(\vn + \Delta),\no
a^\dagger_{\sigma,{\rm NL}}(\vn)&=& a^\dagger_\sigma(\vn)+s_{\rm NL} \sum_{{\bm \Delta} \in S_{\bm \Delta}} a^\dagger_\sigma(\vn + \Delta).
\eea
For the unitary Fermi gas, we set  $C_0=-0.744917$ in lattice units and $s_{\rm NL}=-9.5337 \times 10^{-4}$, which corresponds to infinite scattering length and an effective range of about $0.06$ lattice units.  In Fig.~\ref{phase_shift}, we show the S-wave phase shifts computed on the BCC lattice versus relative momentum.  At low momenta, the phase shifts are in excellent agreement with the unitary limit, which corresponds to a constant phase shift equal to $90$ degrees.  The lattice phase shifts are determined using the spherical wall method \cite{Borasoy:2007vy,Lu:2015riz}.

\begin{center}
\section{Auxiliary-field Monte Carlo simulations}
\end{center}
The transfer matrix $M$ is defined as the normal-ordered time evolution operator,
\begin{equation}
    M = :\exp[-Ha_t]:.
\end{equation}
Let $|\Psi_I \rangle$ and $|\Psi_F \rangle$ be the initial and final state wave functions respectively.  So long as the initial and final states have nonzero overlap with the ground state, we can project out the ground state by multiplying powers of $M$.  The projection amplitude is given by
\begin{equation}
    Z(L_t) = \braket{ \Psi_F | \tilde{M}^{L'_t} M^{L_t} \tilde{M}^{L'_t} | \Psi_I }.
\end{equation}
In order to reduce the number of time steps needed to converge to the ground state, we have multiplied the initial and final states by $L'_t$ powers of the modified transfer matrix,
\begin{equation}
    \tilde{M} = :\exp[-\tilde{H}\tilde{a}_t]:.
\end{equation}
In this modified transfer matrix, the interaction coefficient is stronger, with interaction coefficient $\tilde{C}_0=1.6 C_0$, and the time step $\tilde{a}_t$ is taken to be larger, $\tilde{a}_t=5a_t$.
We compute the ground state energy, $E_0$ by taking the limit 
\bea
E_0 = 
\lim_{L_t \rightarrow \infty} a_t^{-1}\log \frac{Z(L_t-1)}{Z(L_t)}.
\eea 
For the calculations performed here, we take $|\Psi_I \rangle$ and $|\Psi_F \rangle$ to be the same and equal to the Slater determinant state corresponding to the ground state of the free Fermi gas, $| \Psi_0^{\rm free} \rangle$.
For the system we are considering with $33$ spin-up and $33$ spin-down fermions in a periodic cube, the free Fermi ground state is unique.  This corresponds to momentum states with $|{\bm p}|$ less than or equal to $4\pi\hbar/(La_{\rm latt})$.

We can write the interaction at time step $n_t$ and position $\mathbf{n}$ using an auxiliary field $s(\mathbf{n},n_t)$.  We define $V_s^{(n_t)}$ to be
\begin{equation}
    V_s^{(n_t)} = \sqrt{-C_0}\sum_{\mathbf{n}} s(\mathbf{n}, n_t) \rho_{\rm NL}(\mathbf{n}),
\end{equation}
and the quadratic part of the auxiliary-field action is defined as
\begin{equation}
    V_{ss}^{(n_t)} = \frac{1}{2} \sum_{\mathbf{n}}s^2(\mathbf{n}, n_t).
\end{equation}
Then the transfer matrix at time step $n_t$ can be expressed as
\begin{equation}
    M^{(n_t)} = \prod_{\mathbf{n}} \bigg [ \int_{-\infty}^{\infty} \sqrt{\frac{1}{2\pi}}ds(\mathbf{n},n_t) \bigg ] \\
    :\exp[-H_{\rm free} a_t - V_s^{(n_t)}\sqrt{a_t} - V_{ss}^{(n_t)}]:. 
\end{equation}
Instead of the particles interacting with each other, the particles now only interact with the auxiliary field.\\

\section{Results}
We perform BCC lattice simulations of $N_{\uparrow} = 33$ spin-up fermions and $N_{\downarrow} = 33$ spin-down fermions in the unitary limit with lattice lengths
 $L= 4, \cdots , 10$.  This covers a range from $2L^3=128$ lattice points for $L=4$ and $2L^3=2000$ lattice points for $L=10$.  The range is comparable to the $L=5, \cdots, 11$ simple cubic lattices used in Ref.~\cite{He:2019ipt}, with $L^3=125$ points for $L=5$ and $L^3=1331$ points for $L=11$.  For the auxiliary field updates, we use the shuttle algorithm described in Ref.~\cite{Lu:2019nbg}. More details about how the lattice calculations are performed can be found in Ref.~\cite{Lee:2008fa,Lahde:2019npb}. 

As discussed in Ref.~\cite{Bour:2011xt}, there are two different ways to define the ground state ratio $\xi=E_0/E_{\rm FG}$ for a system with a finite number of particles.  One can either define $E_{\rm FG}$ as the ground state energy of the free fermion gas
computed with the same lattice parameters or from the asymptotic formula valid in the thermodynamic and continuum limits.  We define $\xi^{\rm few}$ to be the ratio computed using the free fermion gas energy on the lattice with the same lattice parameters, $E^{\rm few}_{\rm FG}$.  We define $\xi^{\rm thermo}$ to be the ratio computed using $E^{\rm thermo}_{\rm FG}$, where
\bea
E^{\rm thermo}_{\rm FG} = \frac{3}{5}\frac{k_{F}^2}{2m}(N_{\uparrow}+ N_{\downarrow}),
\eea 
where 
\bea
k_F=\frac{(6\pi^2N_{\uparrow})^{1/3}\hbar}{L}=\frac{(6\pi^2N_{\downarrow})^{1/3}\hbar}{L}.
\eea
For $N_{\uparrow} = 33$ and $N_{\downarrow} = 33$, we find that, in the continuum limit, $\xi^{\rm thermo}/\xi^{\rm few}=0.995$.  In the following, we will therefore restrict our focus to computing $\xi^{\rm few}$ and multiply by $0.995$ to obtain $\xi^{\rm thermo}$.

To find the ground state energy, we need to take the limit 
\bea
E_0 = 
\lim_{L_t \rightarrow \infty} E(L_t),
\eea
where
\bea
E(L_t) = a_t^{-1}\log \frac{Z(L_t-1)}{Z(L_t)}.
\eea 
Since we will work with ground state ratios, it is convenient to define
\bea
\xi^{\rm few}(L_t) = \frac{E(L_t)}{E^{\rm few}_{\rm FG}}.
\eea
We perform the extrapolation to an infinite number of time steps by using the asymptotic form
\bea
\xi^{\rm few}(L_t) \simeq \xi^{\rm few} + \beta e^{-L_t a_t \Delta E}\, , \label{asymptotic}
\eea
where the unknown parameters $\xi^{\rm few}, \beta, \Delta E$ are fitted.  When there are multiple sets of data available for the same system, we perform simultaneous fits using the asymptotic form given in Eq.~(\ref{asymptotic}), with a common value for the extrapolated parameter $\xi^{\rm few}$. 

\begin{figure}[h]
\centering
\includegraphics[width=0.48\linewidth]{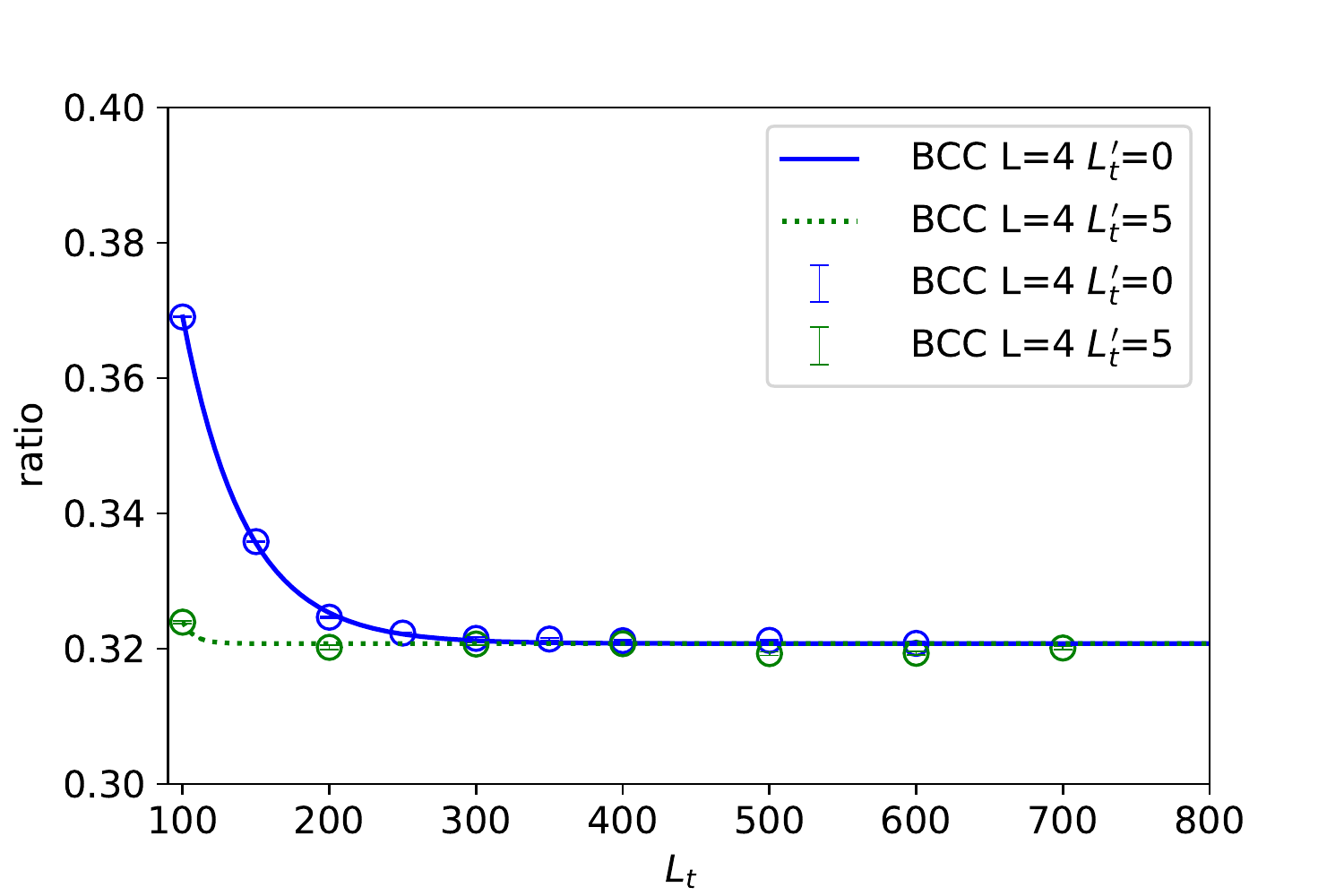}
\includegraphics[width=0.48\linewidth]{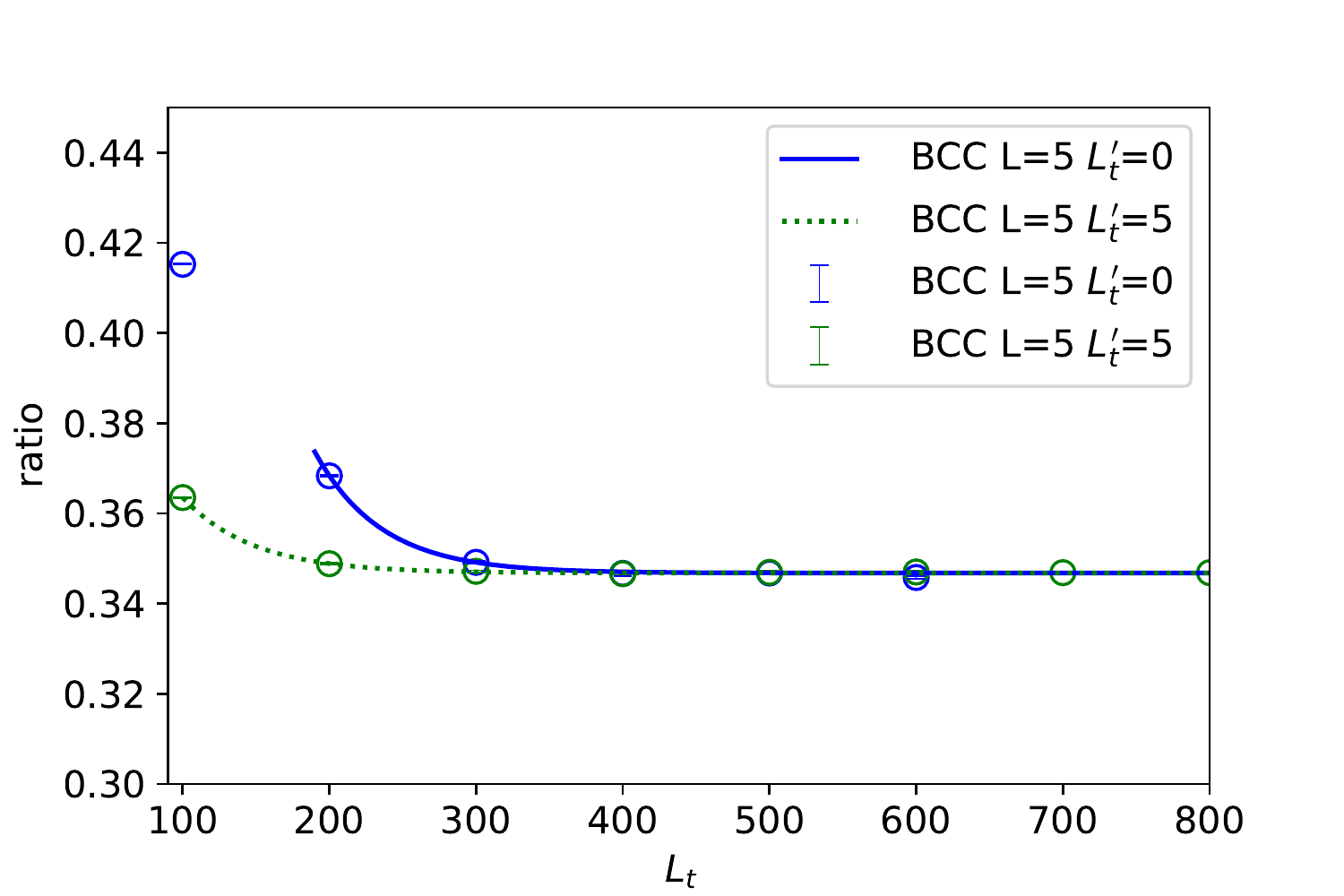}
\caption{Lattice results for $\xi^{\rm few}(L_t)$ versus $L_t$ for the $L=4$ and $L=5$ BCC lattices.  In each case we present data for $L'_t = 0$ and $L'_t=5$ and show the results of the asymptotic fits.} \label{fig:ltfitl45}
\end{figure}

\begin{figure}[h]
\centering
\includegraphics[width=0.48\linewidth]{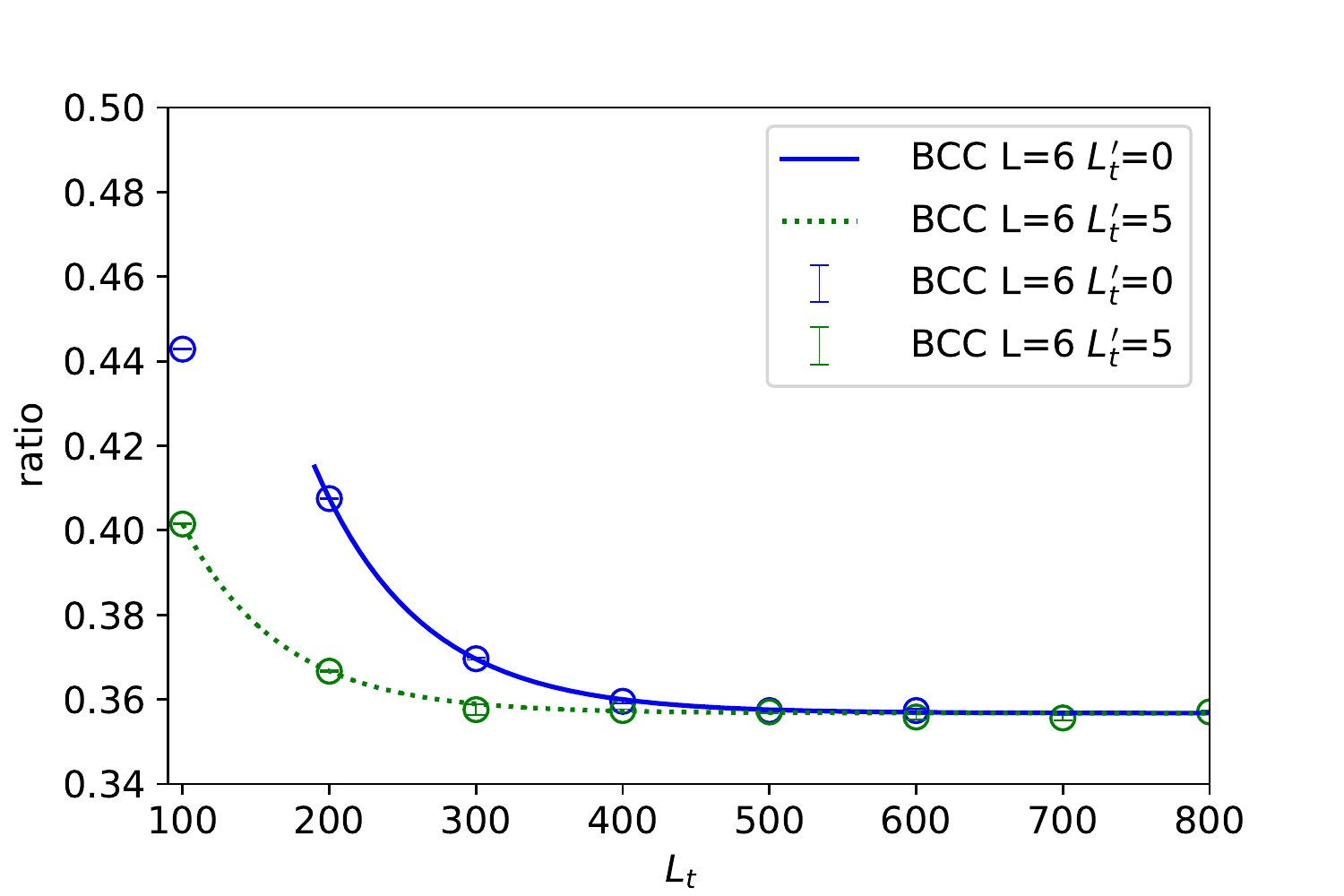}
\includegraphics[width=0.48\linewidth]{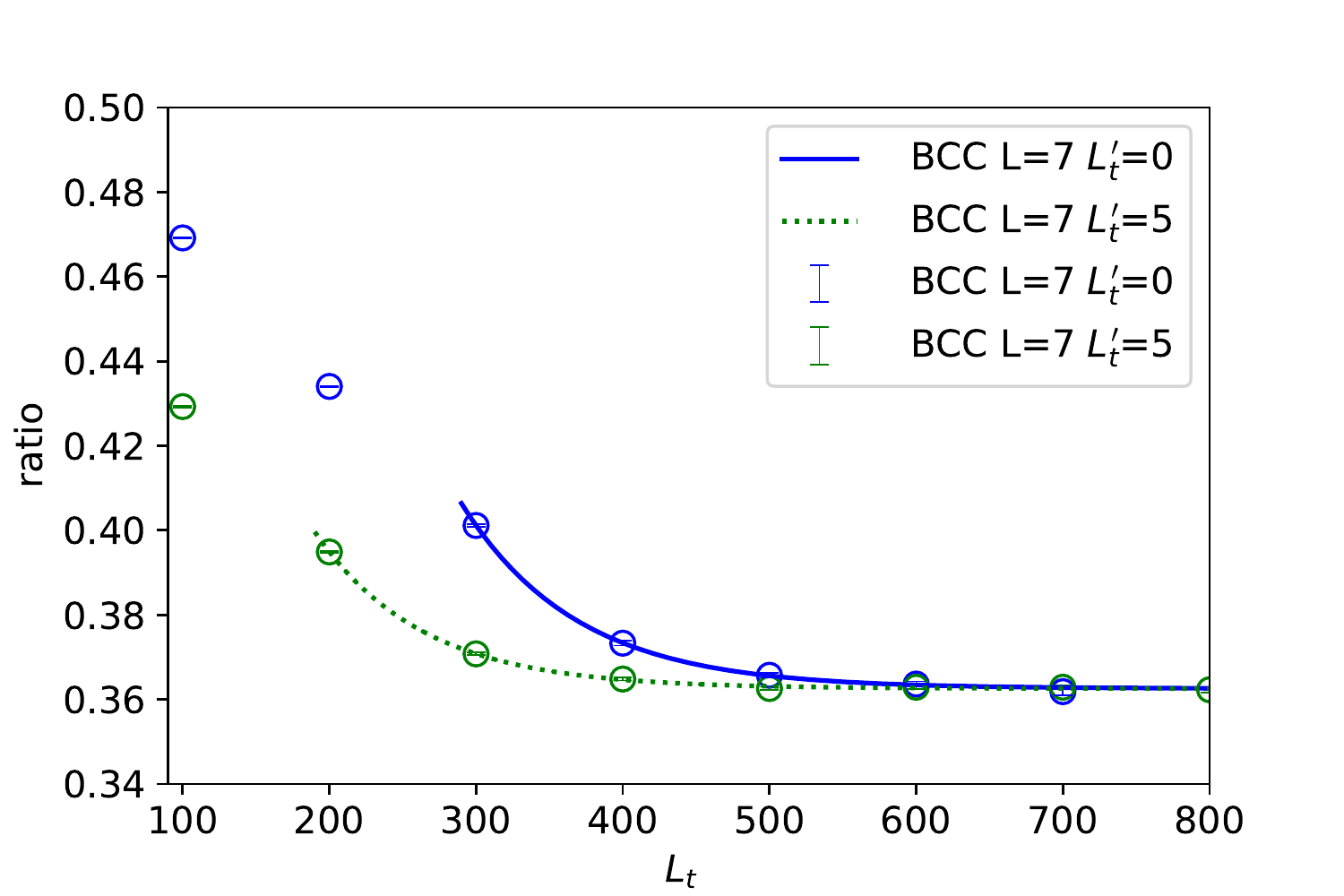}
\caption{Lattice results for $\xi^{\rm few}(L_t)$ versus $L_t$ for the $L=6$ and $L=7$ BCC lattices.  We show data for $L'_t = 0$ and $L'_t=5$ and the results of the asymptotic fits.}	\label{fig:ltfitl67}
\end{figure}

\begin{figure}[h]
\centering
\includegraphics[width=0.48\linewidth]{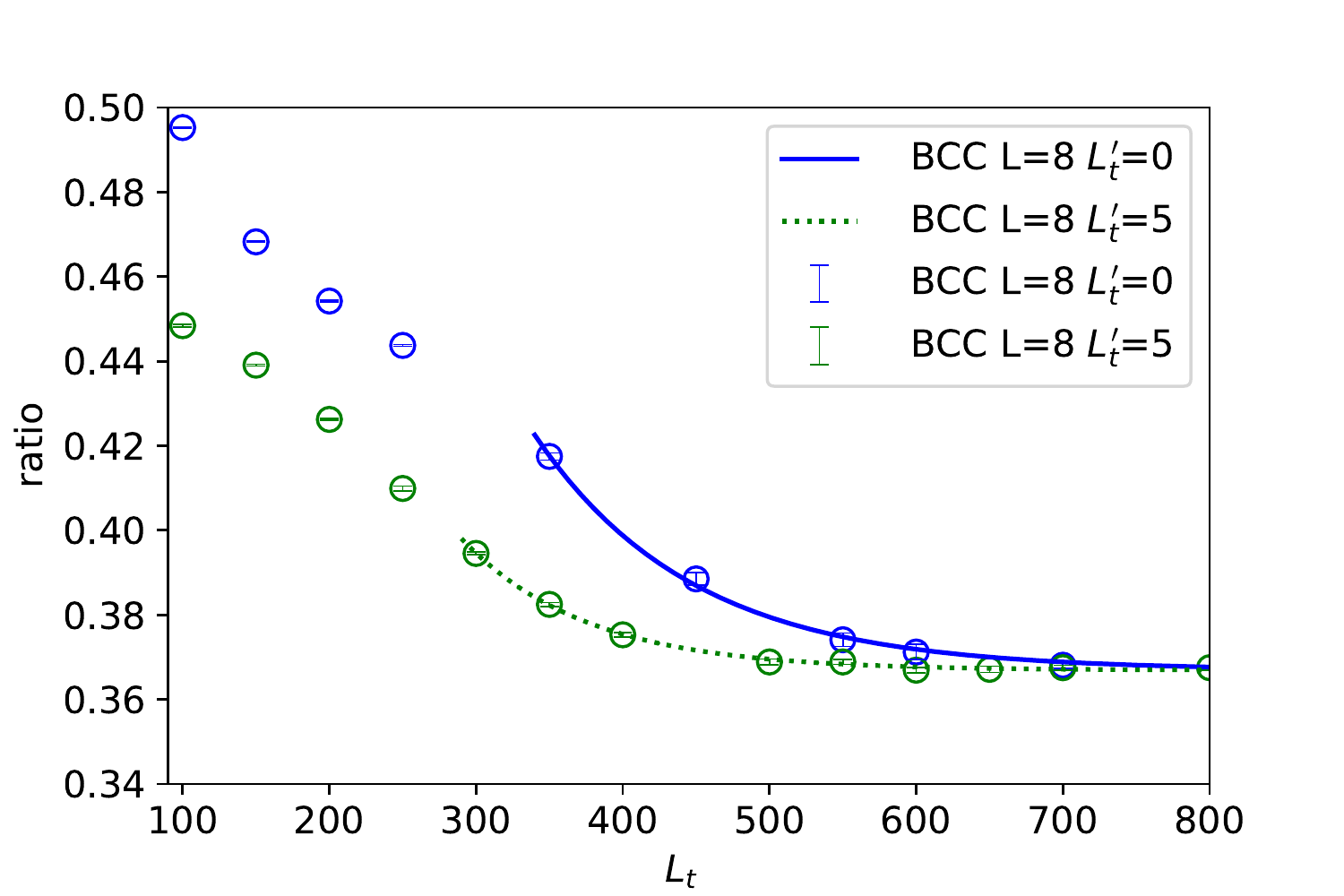}
\includegraphics[width=0.48\linewidth]{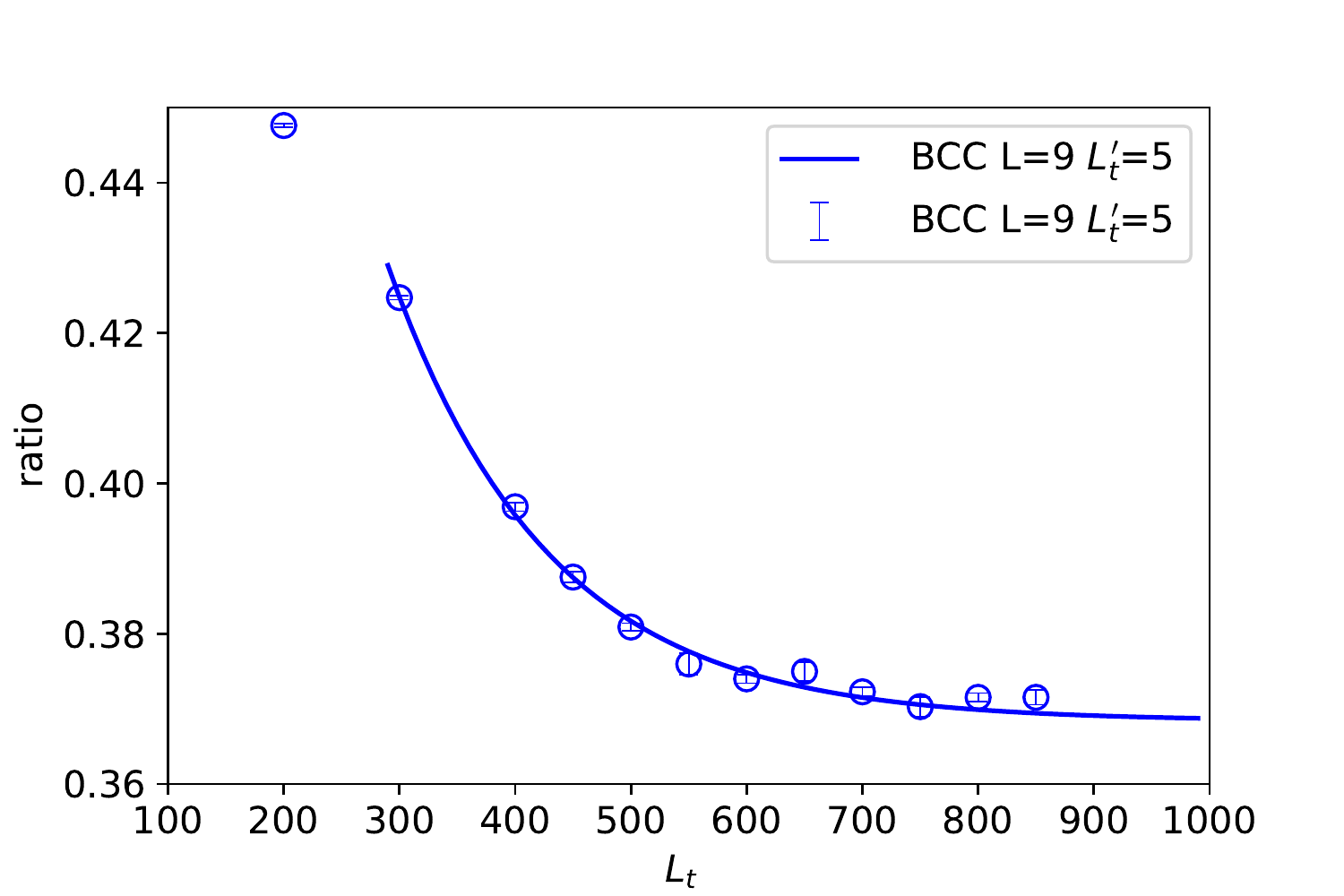}
\caption{Lattice results for $\xi^{\rm few}(L_t)$ versus $L_t$ for the $L=8$ and $L=9$ BCC lattices.  For $L=8$ we show $L'_t = 0$ and $L'_t=5$, while for $L=9$ we present data for $L'_t=5$ only.  In each case we show the results of the asymptotic fits.} \label{fig:ltfitl89}
\end{figure}

\begin{figure}[h]
\centering
\includegraphics[width=0.48\linewidth]{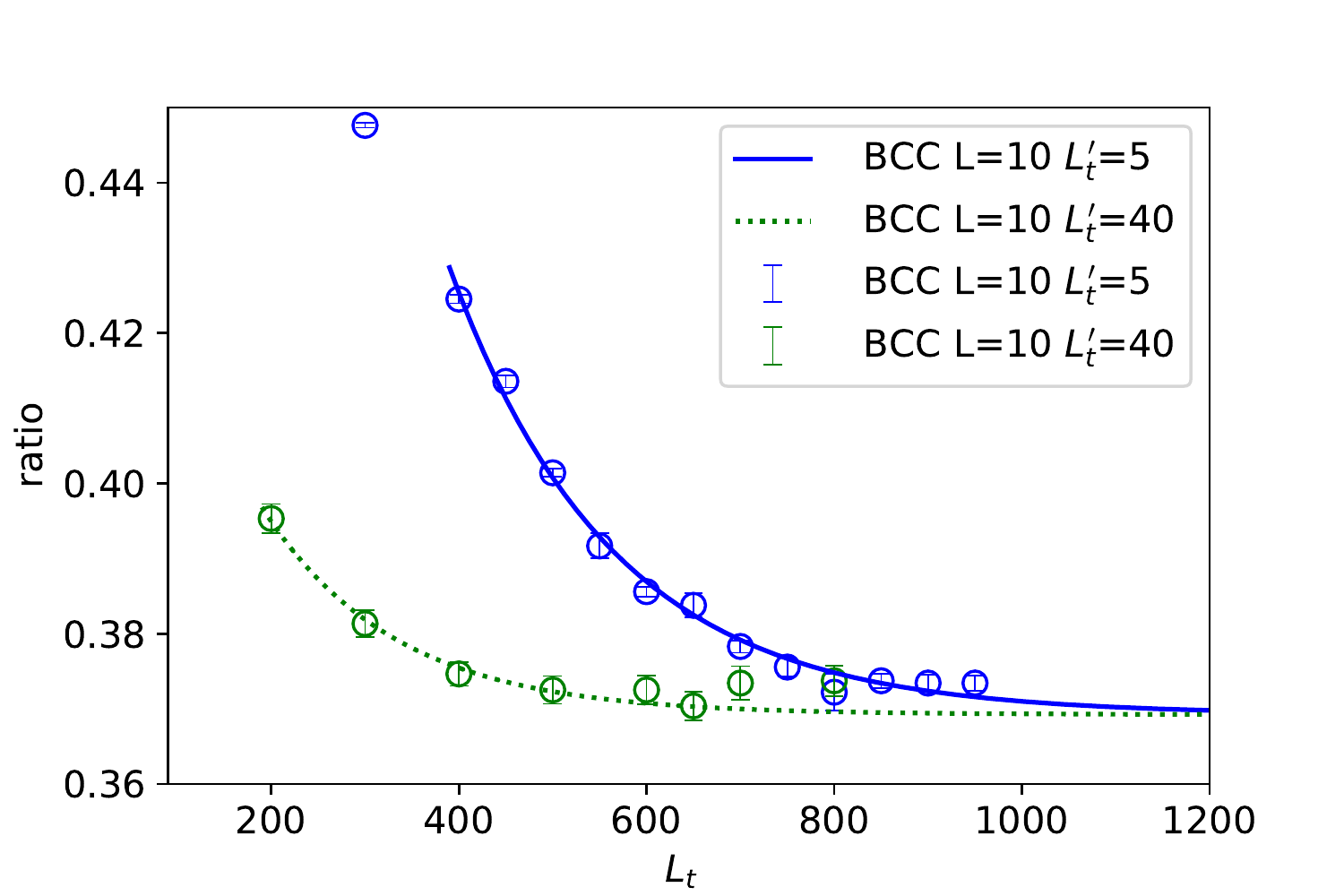}
\caption{Lattice results for $\xi^{\rm few}(L_t)$ versus $L_t$ for the $L=10$ BCC lattice.  We present data for $L'_t = 5$ and $L'_t=40$ and show the results of the asymptotic fits.}\label{fig:ltfitl10}
\end{figure}

In Fig. \ref{fig:ltfitl45}, we plot the lattice results for $\xi^{\rm few}(L_t)$ versus $L_t$ for the $L=4$ and $L=5$ BCC lattices.  In each case we present data for $L'_t = 0$ and $L'_t=5$ and show the results of the asymptotic fits.
In Fig. \ref{fig:ltfitl67}, we plot the lattice results for $\xi^{\rm few}(L_t)$ versus $L_t$ for the $L=6$ and $L=7$ BCC lattices.  We present data for $L'_t = 0$ and $L'_t=5$ and show the results of the asymptotic fits.
In Fig. \ref{fig:ltfitl89}, we plot the lattice results for $\xi^{\rm few}(L_t)$ versus $L_t$ for the $L=8$ and $L=9$ BCC lattices.  For $L=8$ we present data for $L'_t = 0$ and $L'_t=5$, while for $L=9$ we present data for $L'_t=5$ only.  In each case we show the results of the asymptotic fits.
In Fig. \ref{fig:ltfitl10}, we plot the lattice results for $\xi^{\rm few}(L_t)$ versus $L_t$ for the $L=10$ lattice.  We present data for $L'_t = 5$ and $L'_t=40$ and show the results of the asymptotic fits.
We see that all of the asymptotic fits do a good job of reproducing the behavior at large $L_t$.  In Table~\ref{table1}, we show the extracted values for $\xi^{\rm few}(L)$ using BCC lattices for $L=4, \cdots, 10$.

\begin{table}
\caption{\label{table1} Extracted values for $\xi^{\rm few}(L)$ using BCC lattices for $L=4, \cdots, 10$.}
\begin{tabular}{|@{\hspace{1em}}r@{\hspace{1em}}|@{\hspace{1em}}l@{\hspace{1em}}|}
\hline
$L$ & $\xi^{\rm few}(L)$ \\
\hline
4 & 0.3208(2) \\
5 & 0.3468(1) \\
6 & 0.3567(2) \\
7 & 0.3626(2) \\
8 & 0.3669(3) \\
9 & 0.3684(11) \\
10 & 0.3692(12) \\
\hline
\end{tabular}
\end{table}

\begin{figure}[h]
\centering
\includegraphics[width=0.7\linewidth]{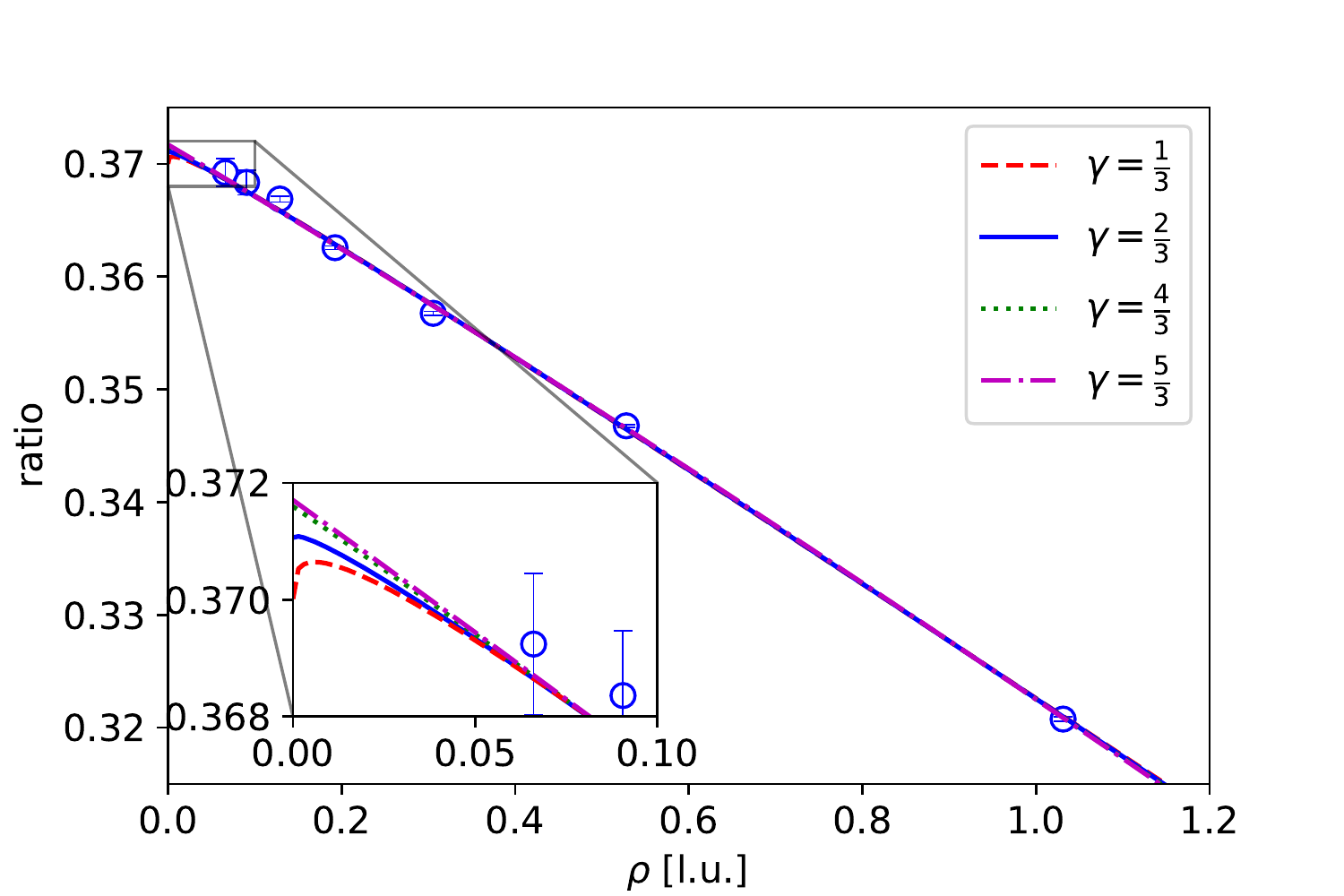}
\caption{Plots of the continuum limit extrapolations of $\xi^{\rm few}(\rho)$ using $\gamma=\frac{1}{3},\frac{2}{3}, \frac{4}{3},\frac{5}{3}$.} \label{fig:rhofit}
\end{figure}

In order to extrapolate $\xi^{\rm few}$ to the continuum limit, we use the fitting function
\bea
\xi^{\rm few}(\rho) = d_1 \rho + d_\gamma \rho^\gamma + \xi^{\rm few},
\eea
where $\rho$ is the density particles in lattice units,
\bea
\rho = \frac{N_{\uparrow}+N_{\downarrow}}{L^3} = \frac{66}{L^3}.
\eea
In Fig. \ref{fig:rhofit} we plot the continuum limit extrapolations for $\xi^{\rm few}(\rho)$ using $\gamma=\frac{1}{3},\frac{2}{3}, \frac{4}{3},\frac{5}{3}$.  We see that the dependence on $\gamma$ is quite minor and the extrapolation is dominated by the $\rho^1$ dependence.  
Incorporating the extrapolation fits for all values of $\gamma$,
we obtain the value $\xi^{\rm few}=0.371(2)$, From this we multiply by $0.995$ to obtain $\xi^{\rm therm}=0.369(2)$.
These are consistent with cubic lattice results $\xi^{\rm finite}=0.372(2)$
and $\xi^{\rm thermo}=0.369(2)$~\cite{He:2019ipt}, as well as other numerical calculations~\cite{Forbes:2010gt, Carlson:2011kv, Endres:2012cw, Jensen:2019zkr} and the experimental measurement of 0.376(4) \cite{Ku:2012}.

\section{Summary and outlook} \label{Sec:Summary}
We have performed quantum many-body calculations using auxiliary-field Monte Carlo simulations on a three-dimensional BCC lattice.  To our knowledge, this is the first calculation of a fermionic many-body system on a BCC lattice. As a benchmark test, we have calculated the ground state energy $E_0$ of 33 spin-up and 33 spin-down fermions in the unitary limit.  Using periodic boxes with lattice lengths $L=4, \cdots, 10$, we find that the ground state energy ratio is $E_0/E_{\rm FG}= 0.369(2)$, $0.371(2)$, using two different definitions of the finite-system energy ratio.  This is in excellent agreement with recent results obtained on a simple cubic lattice \cite{He:2019ipt} using similar auxiliary-field Monte Carlo methods.  The agreement between BCC and simple cubic results gives some confidence that any remnants of the underlying lattice geometry have been removed by performing the continuum limit extrapolation.  In our calculations we have not yet considered the thermodynamic limit where the number of particles are taken to infinity, but we hope to address this challenge in the future.

We find that the amount of computational effort required for the BCC lattice calculations is very similar to that for the simple cubic lattice calculations with the same number of lattice points.  There is a slight increase in computational time arising from the fact that the BCC lattice has eight nearest neighbor sites as opposed to six nearest neighbor sites for the simple cubic lattice.  However, the resulting difference in the computing time is minor.  The quality of the results that one can extract on the BCC lattice seems also quite comparable.  In both cases, we used the shuttle algorithm described in Ref.~\cite{Lu:2019nbg} to update the auxiliary fields. In the lattice calculations presented here, we have not required any Fourier transformation calculations.  Fast Fourier transformations on BCC lattices as well as face-centered cubic (FCC) lattices have been presented in the literature.  Momenta on rectangular lattices are considered in Ref.~\cite{Alim:2009}, while momenta on BCC and FCC lattices are discussed in Ref.~\cite{Zheng:2014}.  

In the future, we hope to present a similar study for lattice calculations on an FCC lattice, which has four sites per unit cell and presents yet another lattice geometry with octahedral symmetry. Having the ability to perform calculations using several different lattice spacings and more than one lattice geometry will be useful for studying the correlations of particles in quantum many-body systems such as the nucleons in an atomic nuclei.  The correlations of nucleons at densities comparable to the saturation density of nuclear matter or greater are of much interest for both atomic nuclei and the equation of state of dense matter in neutron stars.  Obtaining consistent results from simulations using different lattice geometries would mitigate the need to perform computationally expensive calculations with the lattice spacing much smaller than $1$~fm.  

\begin{acknowledgments}
We are grateful for discussions with Ulf-G. Meißner.  Computational resources were partly provided by the National Supercomputing Center of Korea with supercomputing resources including technical support (KSC-2018-COL-0002, KSC-2020-CRE-0027).
The work of Y.H.S. and Y.K. was supported by the Rare Isotope Science Project of Institute for Basic Science, funded by Ministry of Science and ICT (MSICT) and by National Research Foundation of Korea (2013M7A1A1075764). D.L. acknowledges financial support from the U.S. Department of Energy (DE-SC0013365 and DE-SC0021152) and NUCLEI SciDAC-4 collaboration (DE-SC0018083) as well as computational resources from the Oak Ridge Leadership Computing Facility through the INCITE award ``Ab-initio nuclear structure and nuclear reactions'', the J\"ulich Supercomputing Centre at Forschungszentrum J\"ulich, RWTH Aachen, and Michigan State University.
\end{acknowledgments}

\bibliography{References}

\begin{thebibliography}{80}
\expandafter\ifx\csname natexlab\endcsname\relax\def\natexlab#1{#1}\fi
\expandafter\ifx\csname bibnamefont\endcsname\relax
  \def\bibnamefont#1{#1}\fi
\expandafter\ifx\csname bibfnamefont\endcsname\relax
  \def\bibfnamefont#1{#1}\fi
\expandafter\ifx\csname citenamefont\endcsname\relax
  \def\citenamefont#1{#1}\fi
\expandafter\ifx\csname url\endcsname\relax
  \def\url#1{\texttt{#1}}\fi
\expandafter\ifx\csname urlprefix\endcsname\relax\def\urlprefix{URL }\fi
\providecommand{\bibinfo}[2]{#2}
\providecommand{\eprint}[2][]{\url{#2}}

\bibitem[{\citenamefont{He et~al.}(2019)\citenamefont{He, Li, Lu, and
  Lee}}]{He:2019ipt}
\bibinfo{author}{\bibfnamefont{R.}~\bibnamefont{He}},
  \bibinfo{author}{\bibfnamefont{N.}~\bibnamefont{Li}},
  \bibinfo{author}{\bibfnamefont{B.-N.} \bibnamefont{Lu}}, \bibnamefont{and}
  \bibinfo{author}{\bibfnamefont{D.}~\bibnamefont{Lee}} (\bibinfo{year}{2019}),
  \eprint{1910.01257}.

\bibitem[{\citenamefont{Lee et~al.}(2004)\citenamefont{Lee, Borasoy, and
  Sch\"afer}}]{Lee:2004si}
\bibinfo{author}{\bibfnamefont{D.}~\bibnamefont{Lee}},
  \bibinfo{author}{\bibfnamefont{B.}~\bibnamefont{Borasoy}}, \bibnamefont{and}
  \bibinfo{author}{\bibfnamefont{T.}~\bibnamefont{Sch\"afer}},
  \bibinfo{journal}{Phys. Rev. C} \textbf{\bibinfo{volume}{70}},
  \bibinfo{pages}{014007} (\bibinfo{year}{2004}), \eprint{nucl-th/0402072}.

\bibitem[{\citenamefont{Borasoy
  et~al.}(2007{\natexlab{a}})\citenamefont{Borasoy, Epelbaum, Krebs, Lee, and
  Mei{\ss}ner}}]{Borasoy:2006qn}
\bibinfo{author}{\bibfnamefont{B.}~\bibnamefont{Borasoy}},
  \bibinfo{author}{\bibfnamefont{E.}~\bibnamefont{Epelbaum}},
  \bibinfo{author}{\bibfnamefont{H.}~\bibnamefont{Krebs}},
  \bibinfo{author}{\bibfnamefont{D.}~\bibnamefont{Lee}}, \bibnamefont{and}
  \bibinfo{author}{\bibfnamefont{U.-G.} \bibnamefont{Mei{\ss}ner}},
  \bibinfo{journal}{Eur. Phys. J. A} \textbf{\bibinfo{volume}{31}},
  \bibinfo{pages}{105} (\bibinfo{year}{2007}{\natexlab{a}}),
  \eprint{nucl-th/0611087}.

\bibitem[{\citenamefont{Abe and Seki}(2009)}]{Abe:2007fe}
\bibinfo{author}{\bibfnamefont{T.}~\bibnamefont{Abe}} \bibnamefont{and}
  \bibinfo{author}{\bibfnamefont{R.}~\bibnamefont{Seki}},
  \bibinfo{journal}{Phys. Rev. C} \textbf{\bibinfo{volume}{79}},
  \bibinfo{pages}{054002} (\bibinfo{year}{2009}), \eprint{0708.2523}.

\bibitem[{\citenamefont{Lee}(2009)}]{Lee:2008fa}
\bibinfo{author}{\bibfnamefont{D.}~\bibnamefont{Lee}}, \bibinfo{journal}{Prog.
  Part. Nucl. Phys.} \textbf{\bibinfo{volume}{63}}, \bibinfo{pages}{117}
  (\bibinfo{year}{2009}), \eprint{0804.3501}.

\bibitem[{\citenamefont{Drut and Nicholson}(2013)}]{Drut:2012md}
\bibinfo{author}{\bibfnamefont{J.~E.} \bibnamefont{Drut}} \bibnamefont{and}
  \bibinfo{author}{\bibfnamefont{A.~N.} \bibnamefont{Nicholson}},
  \bibinfo{journal}{J. Phys. G} \textbf{\bibinfo{volume}{40}},
  \bibinfo{pages}{043101} (\bibinfo{year}{2013}), \eprint{1208.6556}.

\bibitem[{\citenamefont{Wlaz\l{}owski et~al.}(2014)\citenamefont{Wlaz\l{}owski,
  Holt, Moroz, Bulgac, and Roche}}]{Wlazlowski:2014jna}
\bibinfo{author}{\bibfnamefont{G.}~\bibnamefont{Wlaz\l{}owski}},
  \bibinfo{author}{\bibfnamefont{J.~W.} \bibnamefont{Holt}},
  \bibinfo{author}{\bibfnamefont{S.}~\bibnamefont{Moroz}},
  \bibinfo{author}{\bibfnamefont{A.}~\bibnamefont{Bulgac}}, \bibnamefont{and}
  \bibinfo{author}{\bibfnamefont{K.~J.} \bibnamefont{Roche}},
  \bibinfo{journal}{Phys. Rev. Lett.} \textbf{\bibinfo{volume}{113}},
  \bibinfo{pages}{182503} (\bibinfo{year}{2014}), \eprint{1403.3753}.

\bibitem[{\citenamefont{Elhatisari et~al.}(2015)\citenamefont{Elhatisari, Lee,
  Rupak, Epelbaum, Krebs, Lähde, Luu, and Meißner}}]{Elhatisari:2015iga}
\bibinfo{author}{\bibfnamefont{S.}~\bibnamefont{Elhatisari}},
  \bibinfo{author}{\bibfnamefont{D.}~\bibnamefont{Lee}},
  \bibinfo{author}{\bibfnamefont{G.}~\bibnamefont{Rupak}},
  \bibinfo{author}{\bibfnamefont{E.}~\bibnamefont{Epelbaum}},
  \bibinfo{author}{\bibfnamefont{H.}~\bibnamefont{Krebs}},
  \bibinfo{author}{\bibfnamefont{T.~A.} \bibnamefont{Lähde}},
  \bibinfo{author}{\bibfnamefont{T.}~\bibnamefont{Luu}}, \bibnamefont{and}
  \bibinfo{author}{\bibfnamefont{U.-G.} \bibnamefont{Meißner}},
  \bibinfo{journal}{Nature} \textbf{\bibinfo{volume}{528}},
  \bibinfo{pages}{111} (\bibinfo{year}{2015}), \eprint{1506.03513}.

\bibitem[{\citenamefont{Elhatisari et~al.}(2016)}]{Elhatisari:2016owd}
\bibinfo{author}{\bibfnamefont{S.}~\bibnamefont{Elhatisari}}
  \bibnamefont{et~al.}, \bibinfo{journal}{Phys. Rev. Lett.}
  \textbf{\bibinfo{volume}{117}}, \bibinfo{pages}{132501}
  (\bibinfo{year}{2016}), \eprint{1602.04539}.

\bibitem[{\citenamefont{K\"orber et~al.}(2017)\citenamefont{K\"orber,
  Berkowitz, and Luu}}]{Korber:2017emn}
\bibinfo{author}{\bibfnamefont{C.}~\bibnamefont{K\"orber}},
  \bibinfo{author}{\bibfnamefont{E.}~\bibnamefont{Berkowitz}},
  \bibnamefont{and} \bibinfo{author}{\bibfnamefont{T.}~\bibnamefont{Luu}},
  \bibinfo{journal}{EPL} \textbf{\bibinfo{volume}{119}}, \bibinfo{pages}{60006}
  (\bibinfo{year}{2017}), \eprint{1706.06494}.

\bibitem[{\citenamefont{Elhatisari et~al.}(2017)\citenamefont{Elhatisari,
  Epelbaum, Krebs, Lähde, Lee, Li, Lu, Meißner, and
  Rupak}}]{Elhatisari:2017eno}
\bibinfo{author}{\bibfnamefont{S.}~\bibnamefont{Elhatisari}},
  \bibinfo{author}{\bibfnamefont{E.}~\bibnamefont{Epelbaum}},
  \bibinfo{author}{\bibfnamefont{H.}~\bibnamefont{Krebs}},
  \bibinfo{author}{\bibfnamefont{T.~A.} \bibnamefont{Lähde}},
  \bibinfo{author}{\bibfnamefont{D.}~\bibnamefont{Lee}},
  \bibinfo{author}{\bibfnamefont{N.}~\bibnamefont{Li}},
  \bibinfo{author}{\bibfnamefont{B.-n.} \bibnamefont{Lu}},
  \bibinfo{author}{\bibfnamefont{U.-G.} \bibnamefont{Meißner}},
  \bibnamefont{and} \bibinfo{author}{\bibfnamefont{G.}~\bibnamefont{Rupak}},
  \bibinfo{journal}{Phys. Rev. Lett.} \textbf{\bibinfo{volume}{119}},
  \bibinfo{pages}{222505} (\bibinfo{year}{2017}), \eprint{1702.05177}.

\bibitem[{\citenamefont{Lu et~al.}(2019)\citenamefont{Lu, Li, Elhatisari, Lee,
  Epelbaum, and Meißner}}]{Lu:2018bat}
\bibinfo{author}{\bibfnamefont{B.-N.} \bibnamefont{Lu}},
  \bibinfo{author}{\bibfnamefont{N.}~\bibnamefont{Li}},
  \bibinfo{author}{\bibfnamefont{S.}~\bibnamefont{Elhatisari}},
  \bibinfo{author}{\bibfnamefont{D.}~\bibnamefont{Lee}},
  \bibinfo{author}{\bibfnamefont{E.}~\bibnamefont{Epelbaum}}, \bibnamefont{and}
  \bibinfo{author}{\bibfnamefont{U.-G.} \bibnamefont{Meißner}},
  \bibinfo{journal}{Phys. Lett. B} \textbf{\bibinfo{volume}{797}},
  \bibinfo{pages}{134863} (\bibinfo{year}{2019}), \eprint{1812.10928}.

\bibitem[{\citenamefont{Lähde and Meißner}(2019)}]{Lahde:2019npb}
\bibinfo{author}{\bibfnamefont{T.~A.} \bibnamefont{Lähde}} \bibnamefont{and}
  \bibinfo{author}{\bibfnamefont{U.-G.} \bibnamefont{Meißner}},
  \emph{\bibinfo{title}{{Nuclear Lattice Effective Field Theory}: {An
  introduction}}}, vol. \bibinfo{volume}{957} (\bibinfo{publisher}{Springer},
  \bibinfo{year}{2019}), ISBN \bibinfo{isbn}{978-3-030-14187-5,
  978-3-030-14189-9}.

\bibitem[{\citenamefont{Lu et~al.}(2020)\citenamefont{Lu, Li, Elhatisari, Lee,
  Drut, L\"ahde, Epelbaum, and Mei\ss{}ner}}]{Lu:2019nbg}
\bibinfo{author}{\bibfnamefont{B.-N.} \bibnamefont{Lu}},
  \bibinfo{author}{\bibfnamefont{N.}~\bibnamefont{Li}},
  \bibinfo{author}{\bibfnamefont{S.}~\bibnamefont{Elhatisari}},
  \bibinfo{author}{\bibfnamefont{D.}~\bibnamefont{Lee}},
  \bibinfo{author}{\bibfnamefont{J.~E.} \bibnamefont{Drut}},
  \bibinfo{author}{\bibfnamefont{T.~A.} \bibnamefont{L\"ahde}},
  \bibinfo{author}{\bibfnamefont{E.}~\bibnamefont{Epelbaum}}, \bibnamefont{and}
  \bibinfo{author}{\bibfnamefont{U.-G.} \bibnamefont{Mei\ss{}ner}},
  \bibinfo{journal}{Phys. Rev. Lett.} \textbf{\bibinfo{volume}{125}},
  \bibinfo{pages}{192502} (\bibinfo{year}{2020}), \eprint{1912.05105}.

\bibitem[{\citenamefont{Alexandru et~al.}(2021)\citenamefont{Alexandru,
  Bedaque, Berkowitz, and Warrington}}]{Alexandru:2020zti}
\bibinfo{author}{\bibfnamefont{A.}~\bibnamefont{Alexandru}},
  \bibinfo{author}{\bibfnamefont{P.}~\bibnamefont{Bedaque}},
  \bibinfo{author}{\bibfnamefont{E.}~\bibnamefont{Berkowitz}},
  \bibnamefont{and} \bibinfo{author}{\bibfnamefont{N.~C.}
  \bibnamefont{Warrington}}, \bibinfo{journal}{Phys. Rev. Lett.}
  \textbf{\bibinfo{volume}{126}}, \bibinfo{pages}{132701}
  (\bibinfo{year}{2021}), \eprint{2008.02824}.

\bibitem[{\citenamefont{{Guttman}}(1961)}]{Guttman:1961}
\bibinfo{author}{\bibfnamefont{L.}~\bibnamefont{{Guttman}}},
  \bibinfo{journal}{\jcp} \textbf{\bibinfo{volume}{34}}, \bibinfo{pages}{1024}
  (\bibinfo{year}{1961}).

\bibitem[{\citenamefont{Adler et~al.}(1993)\citenamefont{Adler, Holm, and
  Janke}}]{Adler:1993kp}
\bibinfo{author}{\bibfnamefont{J.}~\bibnamefont{Adler}},
  \bibinfo{author}{\bibfnamefont{C.}~\bibnamefont{Holm}}, \bibnamefont{and}
  \bibinfo{author}{\bibfnamefont{W.}~\bibnamefont{Janke}},
  \bibinfo{journal}{Physica A} \textbf{\bibinfo{volume}{201}},
  \bibinfo{pages}{581} (\bibinfo{year}{1993}), \eprint{hep-lat/9305005}.

\bibitem[{\citenamefont{Caselle et~al.}(1994)\citenamefont{Caselle, Gliozzi,
  and Vinti}}]{Caselle:1993jh}
\bibinfo{author}{\bibfnamefont{M.}~\bibnamefont{Caselle}},
  \bibinfo{author}{\bibfnamefont{F.}~\bibnamefont{Gliozzi}}, \bibnamefont{and}
  \bibinfo{author}{\bibfnamefont{S.}~\bibnamefont{Vinti}},
  \bibinfo{journal}{Nucl. Phys. B Proc. Suppl.} \textbf{\bibinfo{volume}{34}},
  \bibinfo{pages}{726} (\bibinfo{year}{1994}), \eprint{hep-lat/9403006}.

\bibitem[{\citenamefont{Butera and Comi}(1995)}]{Butera:1995kr}
\bibinfo{author}{\bibfnamefont{P.}~\bibnamefont{Butera}} \bibnamefont{and}
  \bibinfo{author}{\bibfnamefont{M.}~\bibnamefont{Comi}},
  \bibinfo{journal}{Phys. Rev. B} \textbf{\bibinfo{volume}{52}},
  \bibinfo{pages}{6185} (\bibinfo{year}{1995}), \eprint{hep-lat/9505027}.

\bibitem[{\citenamefont{Butera and Comi}(1997)}]{Butera:1997ak}
\bibinfo{author}{\bibfnamefont{P.}~\bibnamefont{Butera}} \bibnamefont{and}
  \bibinfo{author}{\bibfnamefont{M.}~\bibnamefont{Comi}},
  \bibinfo{journal}{Phys. Rev. B} \textbf{\bibinfo{volume}{56}},
  \bibinfo{pages}{8212} (\bibinfo{year}{1997}), \eprint{hep-lat/9703018}.

\bibitem[{\citenamefont{Butera and Comi}(1998)}]{Butera:1998rk}
\bibinfo{author}{\bibfnamefont{P.}~\bibnamefont{Butera}} \bibnamefont{and}
  \bibinfo{author}{\bibfnamefont{M.}~\bibnamefont{Comi}},
  \bibinfo{journal}{Phys. Rev. B} \textbf{\bibinfo{volume}{58}},
  \bibinfo{pages}{11552} (\bibinfo{year}{1998}), \eprint{hep-lat/9805025}.

\bibitem[{\citenamefont{Butera and Comi}(1999)}]{Butera:1999qa}
\bibinfo{author}{\bibfnamefont{P.}~\bibnamefont{Butera}} \bibnamefont{and}
  \bibinfo{author}{\bibfnamefont{M.}~\bibnamefont{Comi}},
  \bibinfo{journal}{Phys. Rev. B} \textbf{\bibinfo{volume}{60}},
  \bibinfo{pages}{6749} (\bibinfo{year}{1999}), \eprint{hep-lat/9903010}.

\bibitem[{\citenamefont{Campostrini}(2001)}]{Campostrini:2000yf}
\bibinfo{author}{\bibfnamefont{M.}~\bibnamefont{Campostrini}},
  \bibinfo{journal}{J. Statist. Phys.} \textbf{\bibinfo{volume}{103}},
  \bibinfo{pages}{369} (\bibinfo{year}{2001}), \eprint{cond-mat/0005130}.

\bibitem[{\citenamefont{Butera and Comi}(2000)}]{Butera:2000zt}
\bibinfo{author}{\bibfnamefont{P.}~\bibnamefont{Butera}} \bibnamefont{and}
  \bibinfo{author}{\bibfnamefont{M.}~\bibnamefont{Comi}},
  \bibinfo{journal}{Phys. Rev. B} \textbf{\bibinfo{volume}{62}},
  \bibinfo{pages}{14837} (\bibinfo{year}{2000}), \eprint{hep-lat/0006009}.

\bibitem[{\citenamefont{Butera and Pernici}(2018)}]{Butera:2018kjb}
\bibinfo{author}{\bibfnamefont{P.}~\bibnamefont{Butera}} \bibnamefont{and}
  \bibinfo{author}{\bibfnamefont{M.}~\bibnamefont{Pernici}},
  \bibinfo{journal}{Physica A} \textbf{\bibinfo{volume}{507}},
  \bibinfo{pages}{22} (\bibinfo{year}{2018}), \eprint{1805.04037}.

\bibitem[{\citenamefont{Radi\v{c}evi\'c}(2019)}]{Radicevic:2019vyb}
\bibinfo{author}{\bibfnamefont{D.}~\bibnamefont{Radi\v{c}evi\'c}}
  (\bibinfo{year}{2019}), \eprint{1910.06336}.

\bibitem[{\citenamefont{{Shirley} et~al.}(2019)\citenamefont{{Shirley},
  {Slagle}, and {Chen}}}]{Shirley:2019}
\bibinfo{author}{\bibfnamefont{W.}~\bibnamefont{{Shirley}}},
  \bibinfo{author}{\bibfnamefont{K.}~\bibnamefont{{Slagle}}}, \bibnamefont{and}
  \bibinfo{author}{\bibfnamefont{X.}~\bibnamefont{{Chen}}},
  \bibinfo{journal}{SciPost Physics} \textbf{\bibinfo{volume}{6}},
  \bibinfo{eid}{041} (\bibinfo{year}{2019}), \eprint{1806.08679}.

\bibitem[{\citenamefont{{Murtazaev} et~al.}(2020)\citenamefont{{Murtazaev},
  {Kassan-Ogly}, {Ramazanov}, and {Murtazaev}}}]{Murtazaev:2020}
\bibinfo{author}{\bibfnamefont{A.~K.} \bibnamefont{{Murtazaev}}},
  \bibinfo{author}{\bibfnamefont{F.~A.} \bibnamefont{{Kassan-Ogly}}},
  \bibinfo{author}{\bibfnamefont{M.~K.} \bibnamefont{{Ramazanov}}},
  \bibnamefont{and} \bibinfo{author}{\bibfnamefont{K.~S.}
  \bibnamefont{{Murtazaev}}}, \bibinfo{journal}{Physics of Metals and
  Metallography} \textbf{\bibinfo{volume}{121}}, \bibinfo{pages}{305}
  (\bibinfo{year}{2020}).

\bibitem[{\citenamefont{{You} et~al.}(2020)\citenamefont{{You}, {Devakul},
  {Burnell}, and {Sondhi}}}]{You:2020}
\bibinfo{author}{\bibfnamefont{Y.}~\bibnamefont{{You}}},
  \bibinfo{author}{\bibfnamefont{T.}~\bibnamefont{{Devakul}}},
  \bibinfo{author}{\bibfnamefont{F.~J.} \bibnamefont{{Burnell}}},
  \bibnamefont{and} \bibinfo{author}{\bibfnamefont{S.~L.}
  \bibnamefont{{Sondhi}}}, \bibinfo{journal}{Annals of Physics}
  \textbf{\bibinfo{volume}{416}}, \bibinfo{eid}{168140} (\bibinfo{year}{2020}),
  \eprint{1805.09800}.

\bibitem[{\citenamefont{{O'Hara} et~al.}(2002)\citenamefont{{O'Hara}, {Hemmer},
  {Gehm}, {Granade}, and {Thomas}}}]{O'Hara:2002}
\bibinfo{author}{\bibfnamefont{K.~M.} \bibnamefont{{O'Hara}}},
  \bibinfo{author}{\bibfnamefont{S.~L.} \bibnamefont{{Hemmer}}},
  \bibinfo{author}{\bibfnamefont{M.~E.} \bibnamefont{{Gehm}}},
  \bibinfo{author}{\bibfnamefont{S.~R.} \bibnamefont{{Granade}}},
  \bibnamefont{and} \bibinfo{author}{\bibfnamefont{J.~E.}
  \bibnamefont{{Thomas}}}, \bibinfo{journal}{Science}
  \textbf{\bibinfo{volume}{298}}, \bibinfo{pages}{2179} (\bibinfo{year}{2002}),
  \eprint{cond-mat/0212463}.

\bibitem[{\citenamefont{{Partridge} et~al.}(2006)\citenamefont{{Partridge},
  {Li}, {Kamar}, {Liao}, and {Hulet}}}]{Partridge:2006}
\bibinfo{author}{\bibfnamefont{G.~B.} \bibnamefont{{Partridge}}},
  \bibinfo{author}{\bibfnamefont{W.}~\bibnamefont{{Li}}},
  \bibinfo{author}{\bibfnamefont{R.~I.} \bibnamefont{{Kamar}}},
  \bibinfo{author}{\bibfnamefont{Y.-a.} \bibnamefont{{Liao}}},
  \bibnamefont{and} \bibinfo{author}{\bibfnamefont{R.~G.}
  \bibnamefont{{Hulet}}}, \bibinfo{journal}{Science}
  \textbf{\bibinfo{volume}{311}}, \bibinfo{pages}{503} (\bibinfo{year}{2006}),
  \eprint{cond-mat/0511752}.

\bibitem[{\citenamefont{{Gehm} et~al.}(2003)\citenamefont{{Gehm}, {Hemmer},
  {Granade}, {O'Hara}, and {Thomas}}}]{Gehm:2003}
\bibinfo{author}{\bibfnamefont{M.~E.} \bibnamefont{{Gehm}}},
  \bibinfo{author}{\bibfnamefont{S.~L.} \bibnamefont{{Hemmer}}},
  \bibinfo{author}{\bibfnamefont{S.~R.} \bibnamefont{{Granade}}},
  \bibinfo{author}{\bibfnamefont{K.~M.} \bibnamefont{{O'Hara}}},
  \bibnamefont{and} \bibinfo{author}{\bibfnamefont{J.~E.}
  \bibnamefont{{Thomas}}}, \bibinfo{journal}{\pra}
  \textbf{\bibinfo{volume}{68}}, \bibinfo{eid}{011401} (\bibinfo{year}{2003}),
  \eprint{cond-mat/0212499}.

\bibitem[{\citenamefont{{Bourdel} et~al.}(2004)\citenamefont{{Bourdel},
  {Khaykovich}, {Cubizolles}, {Zhang}, {Chevy}, {Teichmann}, {Tarruell},
  {Kokkelmans}, and {Salomon}}}]{Bourdel:2004}
\bibinfo{author}{\bibfnamefont{T.}~\bibnamefont{{Bourdel}}},
  \bibinfo{author}{\bibfnamefont{L.}~\bibnamefont{{Khaykovich}}},
  \bibinfo{author}{\bibfnamefont{J.}~\bibnamefont{{Cubizolles}}},
  \bibinfo{author}{\bibfnamefont{J.}~\bibnamefont{{Zhang}}},
  \bibinfo{author}{\bibfnamefont{F.}~\bibnamefont{{Chevy}}},
  \bibinfo{author}{\bibfnamefont{M.}~\bibnamefont{{Teichmann}}},
  \bibinfo{author}{\bibfnamefont{L.}~\bibnamefont{{Tarruell}}},
  \bibinfo{author}{\bibfnamefont{S.~J.} \bibnamefont{{Kokkelmans}}},
  \bibnamefont{and}
  \bibinfo{author}{\bibfnamefont{C.}~\bibnamefont{{Salomon}}},
  \bibinfo{journal}{\prl} \textbf{\bibinfo{volume}{93}}, \bibinfo{eid}{050401}
  (\bibinfo{year}{2004}), \eprint{cond-mat/0403091}.

\bibitem[{\citenamefont{{Bartenstein} et~al.}(2004)\citenamefont{{Bartenstein},
  {Altmeyer}, {Riedl}, {Jochim}, {Chin}, {Denschlag}, and
  {Grimm}}}]{Bartenstein:2004}
\bibinfo{author}{\bibfnamefont{M.}~\bibnamefont{{Bartenstein}}},
  \bibinfo{author}{\bibfnamefont{A.}~\bibnamefont{{Altmeyer}}},
  \bibinfo{author}{\bibfnamefont{S.}~\bibnamefont{{Riedl}}},
  \bibinfo{author}{\bibfnamefont{S.}~\bibnamefont{{Jochim}}},
  \bibinfo{author}{\bibfnamefont{C.}~\bibnamefont{{Chin}}},
  \bibinfo{author}{\bibfnamefont{J.~H.} \bibnamefont{{Denschlag}}},
  \bibnamefont{and} \bibinfo{author}{\bibfnamefont{R.}~\bibnamefont{{Grimm}}},
  \bibinfo{journal}{\prl} \textbf{\bibinfo{volume}{92}}, \bibinfo{eid}{120401}
  (\bibinfo{year}{2004}), \eprint{cond-mat/0401109}.

\bibitem[{\citenamefont{{Kinast} et~al.}(2005)\citenamefont{{Kinast},
  {Turlapov}, {Thomas}, {Chen}, {Stajic}, and {Levin}}}]{Kinast:2005}
\bibinfo{author}{\bibfnamefont{J.}~\bibnamefont{{Kinast}}},
  \bibinfo{author}{\bibfnamefont{A.}~\bibnamefont{{Turlapov}}},
  \bibinfo{author}{\bibfnamefont{J.~E.} \bibnamefont{{Thomas}}},
  \bibinfo{author}{\bibfnamefont{Q.}~\bibnamefont{{Chen}}},
  \bibinfo{author}{\bibfnamefont{J.}~\bibnamefont{{Stajic}}}, \bibnamefont{and}
  \bibinfo{author}{\bibfnamefont{K.}~\bibnamefont{{Levin}}},
  \bibinfo{journal}{Science} \textbf{\bibinfo{volume}{307}},
  \bibinfo{pages}{1296} (\bibinfo{year}{2005}), \eprint{cond-mat/0502087}.

\bibitem[{\citenamefont{{Regal} et~al.}(2005)\citenamefont{{Regal}, {Greiner},
  {Giorgini}, {Holland}, and {Jin}}}]{Regal:2005}
\bibinfo{author}{\bibfnamefont{C.~A.} \bibnamefont{{Regal}}},
  \bibinfo{author}{\bibfnamefont{M.}~\bibnamefont{{Greiner}}},
  \bibinfo{author}{\bibfnamefont{S.}~\bibnamefont{{Giorgini}}},
  \bibinfo{author}{\bibfnamefont{M.}~\bibnamefont{{Holland}}},
  \bibnamefont{and} \bibinfo{author}{\bibfnamefont{D.~S.} \bibnamefont{{Jin}}},
  \bibinfo{journal}{\prl} \textbf{\bibinfo{volume}{95}}, \bibinfo{eid}{250404}
  (\bibinfo{year}{2005}), \eprint{cond-mat/0507316}.

\bibitem[{\citenamefont{{Stewart} et~al.}(2006)\citenamefont{{Stewart},
  {Gaebler}, {Regal}, and {Jin}}}]{Stewart:2006}
\bibinfo{author}{\bibfnamefont{J.~T.} \bibnamefont{{Stewart}}},
  \bibinfo{author}{\bibfnamefont{J.~P.} \bibnamefont{{Gaebler}}},
  \bibinfo{author}{\bibfnamefont{C.~A.} \bibnamefont{{Regal}}},
  \bibnamefont{and} \bibinfo{author}{\bibfnamefont{D.~S.} \bibnamefont{{Jin}}},
  \bibinfo{journal}{\prl} \textbf{\bibinfo{volume}{97}}, \bibinfo{eid}{220406}
  (\bibinfo{year}{2006}), \eprint{cond-mat/0607776}.

\bibitem[{\citenamefont{{Luo} and {Thomas}}(2009)}]{Luo:2009}
\bibinfo{author}{\bibfnamefont{L.}~\bibnamefont{{Luo}}} \bibnamefont{and}
  \bibinfo{author}{\bibfnamefont{J.~E.} \bibnamefont{{Thomas}}},
  \bibinfo{journal}{Journal of Low Temperature Physics}
  \textbf{\bibinfo{volume}{154}}, \bibinfo{pages}{1} (\bibinfo{year}{2009}),
  \eprint{0811.1159}.

\bibitem[{\citenamefont{{Navon} et~al.}(2010)\citenamefont{{Navon},
  {Nascimb{\`e}ne}, {Chevy}, and {Salomon}}}]{Navon:2010}
\bibinfo{author}{\bibfnamefont{N.}~\bibnamefont{{Navon}}},
  \bibinfo{author}{\bibfnamefont{S.}~\bibnamefont{{Nascimb{\`e}ne}}},
  \bibinfo{author}{\bibfnamefont{F.}~\bibnamefont{{Chevy}}}, \bibnamefont{and}
  \bibinfo{author}{\bibfnamefont{C.}~\bibnamefont{{Salomon}}},
  \bibinfo{journal}{Science} \textbf{\bibinfo{volume}{328}},
  \bibinfo{pages}{729} (\bibinfo{year}{2010}), \eprint{1004.1465}.

\bibitem[{\citenamefont{{Nascimb{\`e}ne}
  et~al.}(2010)\citenamefont{{Nascimb{\`e}ne}, {Navon}, {Jiang}, {Chevy}, and
  {Salomon}}}]{Nascimbene:2010}
\bibinfo{author}{\bibfnamefont{S.}~\bibnamefont{{Nascimb{\`e}ne}}},
  \bibinfo{author}{\bibfnamefont{N.}~\bibnamefont{{Navon}}},
  \bibinfo{author}{\bibfnamefont{K.~J.} \bibnamefont{{Jiang}}},
  \bibinfo{author}{\bibfnamefont{F.}~\bibnamefont{{Chevy}}}, \bibnamefont{and}
  \bibinfo{author}{\bibfnamefont{C.}~\bibnamefont{{Salomon}}},
  \bibinfo{journal}{\nat} \textbf{\bibinfo{volume}{463}}, \bibinfo{pages}{1057}
  (\bibinfo{year}{2010}), \eprint{0911.0747}.

\bibitem[{\citenamefont{{Ku} et~al.}(2012)\citenamefont{{Ku}, {Sommer},
  {Cheuk}, and {Zwierlein}}}]{Ku:2012}
\bibinfo{author}{\bibfnamefont{M.~J.~H.} \bibnamefont{{Ku}}},
  \bibinfo{author}{\bibfnamefont{A.~T.} \bibnamefont{{Sommer}}},
  \bibinfo{author}{\bibfnamefont{L.~W.} \bibnamefont{{Cheuk}}},
  \bibnamefont{and} \bibinfo{author}{\bibfnamefont{M.~W.}
  \bibnamefont{{Zwierlein}}}, \bibinfo{journal}{Science}
  \textbf{\bibinfo{volume}{335}}, \bibinfo{pages}{563} (\bibinfo{year}{2012}),
  \eprint{1110.3309}.

\bibitem[{\citenamefont{Engelbrecht et~al.}(1997)\citenamefont{Engelbrecht,
  Randeria, and S\'ade~Melo}}]{Engelbrecht:1997}
\bibinfo{author}{\bibfnamefont{J.~R.} \bibnamefont{Engelbrecht}},
  \bibinfo{author}{\bibfnamefont{M.}~\bibnamefont{Randeria}}, \bibnamefont{and}
  \bibinfo{author}{\bibfnamefont{C.~A.~R.} \bibnamefont{S\'ade~Melo}},
  \bibinfo{journal}{Phys. Rev. B} \textbf{\bibinfo{volume}{55}},
  \bibinfo{pages}{15153} (\bibinfo{year}{1997}).

\bibitem[{\citenamefont{{Baker}}(1999)}]{Baker:1999}
\bibinfo{author}{\bibfnamefont{J.}~\bibnamefont{{Baker}},
  \bibfnamefont{George~A.}}, \bibinfo{journal}{\prc}
  \textbf{\bibinfo{volume}{60}}, \bibinfo{eid}{054311} (\bibinfo{year}{1999}).

\bibitem[{\citenamefont{{Steele}}(2000)}]{Steele:2000}
\bibinfo{author}{\bibfnamefont{J.~V.} \bibnamefont{{Steele}}},
  \bibinfo{journal}{arXiv e-prints} \bibinfo{eid}{nucl-th/0010066}
  (\bibinfo{year}{2000}), \eprint{nucl-th/0010066}.

\bibitem[{\citenamefont{{Heiselberg}}(2001)}]{Heiselberg:2001}
\bibinfo{author}{\bibfnamefont{H.}~\bibnamefont{{Heiselberg}}},
  \bibinfo{journal}{\pra} \textbf{\bibinfo{volume}{63}}, \bibinfo{eid}{043606}
  (\bibinfo{year}{2001}), \eprint{cond-mat/0002056}.

\bibitem[{\citenamefont{{Perali} et~al.}(2004)\citenamefont{{Perali}, {Pieri},
  and {Strinati}}}]{Strinati:2004}
\bibinfo{author}{\bibfnamefont{A.}~\bibnamefont{{Perali}}},
  \bibinfo{author}{\bibfnamefont{P.}~\bibnamefont{{Pieri}}}, \bibnamefont{and}
  \bibinfo{author}{\bibfnamefont{G.~C.} \bibnamefont{{Strinati}}},
  \bibinfo{journal}{\prl} \textbf{\bibinfo{volume}{93}}, \bibinfo{eid}{100404}
  (\bibinfo{year}{2004}), \eprint{cond-mat/0405102}.

\bibitem[{\citenamefont{{Sch{\"a}fer} et~al.}(2005)\citenamefont{{Sch{\"a}fer},
  {Kao}, and {Cotanch}}}]{Schafer:2005}
\bibinfo{author}{\bibfnamefont{T.}~\bibnamefont{{Sch{\"a}fer}}},
  \bibinfo{author}{\bibfnamefont{C.~W.} \bibnamefont{{Kao}}}, \bibnamefont{and}
  \bibinfo{author}{\bibfnamefont{S.~R.} \bibnamefont{{Cotanch}}},
  \bibinfo{journal}{Nucl. Phys. A} \textbf{\bibinfo{volume}{762}},
  \bibinfo{pages}{82} (\bibinfo{year}{2005}), \eprint{nucl-th/0504088}.

\bibitem[{\citenamefont{{Papenbrock}}(2005)}]{Papenbrock:2005}
\bibinfo{author}{\bibfnamefont{T.}~\bibnamefont{{Papenbrock}}},
  \bibinfo{journal}{\pra} \textbf{\bibinfo{volume}{72}}, \bibinfo{eid}{041603}
  (\bibinfo{year}{2005}), \eprint{cond-mat/0507183}.

\bibitem[{\citenamefont{{Nishida} and {Son}}(2006)}]{Nishida:2006}
\bibinfo{author}{\bibfnamefont{Y.}~\bibnamefont{{Nishida}}} \bibnamefont{and}
  \bibinfo{author}{\bibfnamefont{D.~T.} \bibnamefont{{Son}}},
  \bibinfo{journal}{\prl} \textbf{\bibinfo{volume}{97}}, \bibinfo{eid}{050403}
  (\bibinfo{year}{2006}), \eprint{cond-mat/0604500}.

\bibitem[{\citenamefont{{Haussmann} et~al.}(2007)\citenamefont{{Haussmann},
  {Rantner}, {Cerrito}, and {Zwerger}}}]{Haussman:2007}
\bibinfo{author}{\bibfnamefont{R.}~\bibnamefont{{Haussmann}}},
  \bibinfo{author}{\bibfnamefont{W.}~\bibnamefont{{Rantner}}},
  \bibinfo{author}{\bibfnamefont{S.}~\bibnamefont{{Cerrito}}},
  \bibnamefont{and}
  \bibinfo{author}{\bibfnamefont{W.}~\bibnamefont{{Zwerger}}},
  \bibinfo{journal}{\pra} \textbf{\bibinfo{volume}{75}}, \bibinfo{eid}{023610}
  (\bibinfo{year}{2007}), \eprint{cond-mat/0608282}.

\bibitem[{\citenamefont{{Veillette} et~al.}(2007)\citenamefont{{Veillette},
  {Sheehy}, and {Radzihovsky}}}]{Veillette:2007}
\bibinfo{author}{\bibfnamefont{M.~Y.} \bibnamefont{{Veillette}}},
  \bibinfo{author}{\bibfnamefont{D.~E.} \bibnamefont{{Sheehy}}},
  \bibnamefont{and}
  \bibinfo{author}{\bibfnamefont{L.}~\bibnamefont{{Radzihovsky}}},
  \bibinfo{journal}{\pra} \textbf{\bibinfo{volume}{75}}, \bibinfo{eid}{043614}
  (\bibinfo{year}{2007}), \eprint{cond-mat/0610798}.

\bibitem[{\citenamefont{{Arnold} et~al.}(2007)\citenamefont{{Arnold}, {Drut},
  and {Son}}}]{Arnold:2007}
\bibinfo{author}{\bibfnamefont{P.}~\bibnamefont{{Arnold}}},
  \bibinfo{author}{\bibfnamefont{J.~E.} \bibnamefont{{Drut}}},
  \bibnamefont{and} \bibinfo{author}{\bibfnamefont{D.~T.} \bibnamefont{{Son}}},
  \bibinfo{journal}{\pra} \textbf{\bibinfo{volume}{75}}, \bibinfo{eid}{043605}
  (\bibinfo{year}{2007}), \eprint{cond-mat/0608477}.

\bibitem[{\citenamefont{Ji-Sheng}(2007)}]{Chen:2007}
\bibinfo{author}{\bibfnamefont{C.}~\bibnamefont{Ji-Sheng}},
  \bibinfo{journal}{Chinese Physics Letters} \textbf{\bibinfo{volume}{24}},
  \bibinfo{eid}{1825-1828} (\bibinfo{year}{2007}).

\bibitem[{\citenamefont{{Nishida}}(2009)}]{Nishida:2009}
\bibinfo{author}{\bibfnamefont{Y.}~\bibnamefont{{Nishida}}},
  \bibinfo{journal}{\pra} \textbf{\bibinfo{volume}{79}}, \bibinfo{eid}{013627}
  (\bibinfo{year}{2009}), \eprint{0808.3826}.

\bibitem[{\citenamefont{Endres et~al.}(2013)\citenamefont{Endres, Kaplan, Lee,
  and Nicholson}}]{Endres:2012cw}
\bibinfo{author}{\bibfnamefont{M.~G.} \bibnamefont{Endres}},
  \bibinfo{author}{\bibfnamefont{D.~B.} \bibnamefont{Kaplan}},
  \bibinfo{author}{\bibfnamefont{J.-W.} \bibnamefont{Lee}}, \bibnamefont{and}
  \bibinfo{author}{\bibfnamefont{A.~N.} \bibnamefont{Nicholson}},
  \bibinfo{journal}{Phys. Rev. A} \textbf{\bibinfo{volume}{87}},
  \bibinfo{pages}{023615} (\bibinfo{year}{2013}), \eprint{1203.3169}.

\bibitem[{\citenamefont{Jensen et~al.}(2020)\citenamefont{Jensen, Gilbreth, and
  Alhassid}}]{Jensen:2019zkr}
\bibinfo{author}{\bibfnamefont{S.}~\bibnamefont{Jensen}},
  \bibinfo{author}{\bibfnamefont{C.~N.} \bibnamefont{Gilbreth}},
  \bibnamefont{and} \bibinfo{author}{\bibfnamefont{Y.}~\bibnamefont{Alhassid}},
  \bibinfo{journal}{Phys. Rev. Lett.} \textbf{\bibinfo{volume}{125}},
  \bibinfo{pages}{043402} (\bibinfo{year}{2020}), \eprint{1906.10117}.

\bibitem[{\citenamefont{Mihaila et~al.}(2011)\citenamefont{Mihaila, Dawson,
  Cooper, Chien, and Timmermans}}]{Mihaila:2011pq}
\bibinfo{author}{\bibfnamefont{B.}~\bibnamefont{Mihaila}},
  \bibinfo{author}{\bibfnamefont{J.~F.} \bibnamefont{Dawson}},
  \bibinfo{author}{\bibfnamefont{F.}~\bibnamefont{Cooper}},
  \bibinfo{author}{\bibfnamefont{C.-C.} \bibnamefont{Chien}}, \bibnamefont{and}
  \bibinfo{author}{\bibfnamefont{E.}~\bibnamefont{Timmermans}},
  \bibinfo{journal}{Phys. Rev. A} \textbf{\bibinfo{volume}{83}},
  \bibinfo{pages}{053637} (\bibinfo{year}{2011}), \eprint{1105.4933}.

\bibitem[{\citenamefont{{Carlson} et~al.}(2003)\citenamefont{{Carlson},
  {Chang}, {Pandharipande}, and {Schmidt}}}]{Carlson:2003}
\bibinfo{author}{\bibfnamefont{J.}~\bibnamefont{{Carlson}}},
  \bibinfo{author}{\bibfnamefont{S.~Y.} \bibnamefont{{Chang}}},
  \bibinfo{author}{\bibfnamefont{V.}~\bibnamefont{{Pandharipande}}},
  \bibnamefont{and}
  \bibinfo{author}{\bibfnamefont{K.}~\bibnamefont{{Schmidt}}},
  \bibinfo{journal}{\prl} \textbf{\bibinfo{volume}{91}}, \bibinfo{eid}{050401}
  (\bibinfo{year}{2003}), \eprint{physics/0303094}.

\bibitem[{\citenamefont{{Chang} et~al.}(2004)\citenamefont{{Chang},
  {Pandharipande}, {Carlson}, and {Schmidt}}}]{Chang:2004}
\bibinfo{author}{\bibfnamefont{S.~Y.} \bibnamefont{{Chang}}},
  \bibinfo{author}{\bibfnamefont{V.~R.} \bibnamefont{{Pandharipande}}},
  \bibinfo{author}{\bibfnamefont{J.}~\bibnamefont{{Carlson}}},
  \bibnamefont{and} \bibinfo{author}{\bibfnamefont{K.~E.}
  \bibnamefont{{Schmidt}}}, \bibinfo{journal}{\pra}
  \textbf{\bibinfo{volume}{70}}, \bibinfo{eid}{043602} (\bibinfo{year}{2004}),
  \eprint{physics/0404115}.

\bibitem[{\citenamefont{{Astrakharchik}
  et~al.}(2004)\citenamefont{{Astrakharchik}, {Boronat}, {Casulleras},
  {Giorgini}, and {S.}}}]{Astrakharchik:2004}
\bibinfo{author}{\bibfnamefont{G.~E.} \bibnamefont{{Astrakharchik}}},
  \bibinfo{author}{\bibfnamefont{J.}~\bibnamefont{{Boronat}}},
  \bibinfo{author}{\bibfnamefont{J.}~\bibnamefont{{Casulleras}}},
  \bibinfo{author}{\bibnamefont{{Giorgini}}}, \bibnamefont{and}
  \bibinfo{author}{\bibnamefont{{S.}}}, \bibinfo{journal}{\prl}
  \textbf{\bibinfo{volume}{93}}, \bibinfo{eid}{200404} (\bibinfo{year}{2004}),
  \eprint{cond-mat/0406113}.

\bibitem[{\citenamefont{{Carlson} and {Reddy}}(2005)}]{Carlson:2005}
\bibinfo{author}{\bibfnamefont{J.}~\bibnamefont{{Carlson}}} \bibnamefont{and}
  \bibinfo{author}{\bibfnamefont{S.}~\bibnamefont{{Reddy}}},
  \bibinfo{journal}{\prl} \textbf{\bibinfo{volume}{95}}, \bibinfo{eid}{060401}
  (\bibinfo{year}{2005}), \eprint{cond-mat/0503256}.

\bibitem[{\citenamefont{{Lee} and
  {Sch{\"a}fer}}(2006{\natexlab{a}})}]{Lee:2006a}
\bibinfo{author}{\bibfnamefont{D.}~\bibnamefont{{Lee}}} \bibnamefont{and}
  \bibinfo{author}{\bibfnamefont{T.}~\bibnamefont{{Sch{\"a}fer}}},
  \bibinfo{journal}{\prc} \textbf{\bibinfo{volume}{73}}, \bibinfo{eid}{015201}
  (\bibinfo{year}{2006}{\natexlab{a}}), \eprint{nucl-th/0509017}.

\bibitem[{\citenamefont{{Lee} and
  {Sch{\"a}fer}}(2006{\natexlab{b}})}]{Lee:2006b}
\bibinfo{author}{\bibfnamefont{D.}~\bibnamefont{{Lee}}} \bibnamefont{and}
  \bibinfo{author}{\bibfnamefont{T.}~\bibnamefont{{Sch{\"a}fer}}},
  \bibinfo{journal}{\prc} \textbf{\bibinfo{volume}{73}}, \bibinfo{eid}{015202}
  (\bibinfo{year}{2006}{\natexlab{b}}), \eprint{nucl-th/0509018}.

\bibitem[{\citenamefont{{Bulgac} et~al.}(2006)\citenamefont{{Bulgac}, {Drut},
  and {Magierski}}}]{Bulgac:2006}
\bibinfo{author}{\bibfnamefont{A.}~\bibnamefont{{Bulgac}}},
  \bibinfo{author}{\bibfnamefont{J.~E.} \bibnamefont{{Drut}}},
  \bibnamefont{and}
  \bibinfo{author}{\bibfnamefont{P.}~\bibnamefont{{Magierski}}},
  \bibinfo{journal}{\prl} \textbf{\bibinfo{volume}{96}}, \bibinfo{eid}{090404}
  (\bibinfo{year}{2006}), \eprint{cond-mat/0505374}.

\bibitem[{\citenamefont{{Lee}}(2006)}]{Lee:2006c}
\bibinfo{author}{\bibfnamefont{D.}~\bibnamefont{{Lee}}},
  \bibinfo{journal}{\prb} \textbf{\bibinfo{volume}{73}}, \bibinfo{eid}{115112}
  (\bibinfo{year}{2006}), \eprint{cond-mat/0511332}.

\bibitem[{\citenamefont{{Juillet}}(2007)}]{Juillet:2007}
\bibinfo{author}{\bibfnamefont{O.}~\bibnamefont{{Juillet}}},
  \bibinfo{journal}{New Journal of Physics} \textbf{\bibinfo{volume}{9}},
  \bibinfo{pages}{163} (\bibinfo{year}{2007}), \eprint{cond-mat/0609063}.

\bibitem[{\citenamefont{{Lee}}(2008{\natexlab{a}})}]{Lee:2008a}
\bibinfo{author}{\bibfnamefont{D.}~\bibnamefont{{Lee}}},
  \bibinfo{journal}{European Physical Journal A} \textbf{\bibinfo{volume}{35}},
  \bibinfo{pages}{171} (\bibinfo{year}{2008}{\natexlab{a}}),
  \eprint{0704.3439}.

\bibitem[{\citenamefont{{Lee}}(2008{\natexlab{b}})}]{Lee:2008b}
\bibinfo{author}{\bibfnamefont{D.}~\bibnamefont{{Lee}}},
  \bibinfo{journal}{\prc} \textbf{\bibinfo{volume}{78}}, \bibinfo{eid}{024001}
  (\bibinfo{year}{2008}{\natexlab{b}}), \eprint{0803.1280}.

\bibitem[{\citenamefont{{Li} et~al.}(2011)\citenamefont{{Li}, {Koloren{\v{c}}},
  and {Mitas}}}]{Li:2011}
\bibinfo{author}{\bibfnamefont{X.}~\bibnamefont{{Li}}},
  \bibinfo{author}{\bibfnamefont{J.}~\bibnamefont{{Koloren{\v{c}}}}},
  \bibnamefont{and} \bibinfo{author}{\bibfnamefont{L.}~\bibnamefont{{Mitas}}},
  \bibinfo{journal}{\pra} \textbf{\bibinfo{volume}{84}}, \bibinfo{eid}{023615}
  (\bibinfo{year}{2011}), \eprint{1105.1748}.

\bibitem[{\citenamefont{{Bulgac} et~al.}(2008)\citenamefont{{Bulgac}, {Drut},
  and {Magierski}}}]{Bulgac:2008}
\bibinfo{author}{\bibfnamefont{A.}~\bibnamefont{{Bulgac}}},
  \bibinfo{author}{\bibfnamefont{J.~E.} \bibnamefont{{Drut}}},
  \bibnamefont{and}
  \bibinfo{author}{\bibfnamefont{P.}~\bibnamefont{{Magierski}}},
  \bibinfo{journal}{\pra} \textbf{\bibinfo{volume}{78}}, \bibinfo{eid}{023625}
  (\bibinfo{year}{2008}), \eprint{0803.3238}.

\bibitem[{\citenamefont{{Abe} and {Seki}}(2009)}]{Abe:2009}
\bibinfo{author}{\bibfnamefont{T.}~\bibnamefont{{Abe}}} \bibnamefont{and}
  \bibinfo{author}{\bibfnamefont{R.}~\bibnamefont{{Seki}}},
  \bibinfo{journal}{\prc} \textbf{\bibinfo{volume}{79}}, \bibinfo{eid}{054003}
  (\bibinfo{year}{2009}), \eprint{0708.2524}.

\bibitem[{\citenamefont{Magierski et~al.}(2009)\citenamefont{Magierski,
  Wlazlowski, Bulgac, and Drut}}]{Magierski:2008wa}
\bibinfo{author}{\bibfnamefont{P.}~\bibnamefont{Magierski}},
  \bibinfo{author}{\bibfnamefont{G.}~\bibnamefont{Wlazlowski}},
  \bibinfo{author}{\bibfnamefont{A.}~\bibnamefont{Bulgac}}, \bibnamefont{and}
  \bibinfo{author}{\bibfnamefont{J.~E.} \bibnamefont{Drut}},
  \bibinfo{journal}{Phys. Rev. Lett.} \textbf{\bibinfo{volume}{103}},
  \bibinfo{pages}{210403} (\bibinfo{year}{2009}), \eprint{0801.1504}.

\bibitem[{\citenamefont{{Gandolfi} et~al.}(2011)\citenamefont{{Gandolfi},
  {Schmidt}, and {Carlson}}}]{Gandolfi:2011}
\bibinfo{author}{\bibfnamefont{S.}~\bibnamefont{{Gandolfi}}},
  \bibinfo{author}{\bibfnamefont{K.~E.} \bibnamefont{{Schmidt}}},
  \bibnamefont{and}
  \bibinfo{author}{\bibfnamefont{J.}~\bibnamefont{{Carlson}}},
  \bibinfo{journal}{\pra} \textbf{\bibinfo{volume}{83}}, \bibinfo{eid}{041601}
  (\bibinfo{year}{2011}), \eprint{1012.4417}.

\bibitem[{\citenamefont{Forbes et~al.}(2011)\citenamefont{Forbes, Gandolfi, and
  Gezerlis}}]{Forbes:2010gt}
\bibinfo{author}{\bibfnamefont{M.~M.} \bibnamefont{Forbes}},
  \bibinfo{author}{\bibfnamefont{S.}~\bibnamefont{Gandolfi}}, \bibnamefont{and}
  \bibinfo{author}{\bibfnamefont{A.}~\bibnamefont{Gezerlis}},
  \bibinfo{journal}{Phys. Rev. Lett.} \textbf{\bibinfo{volume}{106}},
  \bibinfo{pages}{235303} (\bibinfo{year}{2011}), \eprint{1011.2197}.

\bibitem[{\citenamefont{Carlson et~al.}(2011)\citenamefont{Carlson, Gandolfi,
  Schmidt, and Zhang}}]{Carlson:2011kv}
\bibinfo{author}{\bibfnamefont{J.}~\bibnamefont{Carlson}},
  \bibinfo{author}{\bibfnamefont{S.}~\bibnamefont{Gandolfi}},
  \bibinfo{author}{\bibfnamefont{K.~E.} \bibnamefont{Schmidt}},
  \bibnamefont{and} \bibinfo{author}{\bibfnamefont{S.}~\bibnamefont{Zhang}},
  \bibinfo{journal}{Phys. Rev. A} \textbf{\bibinfo{volume}{84}},
  \bibinfo{pages}{061602} (\bibinfo{year}{2011}), \eprint{1107.5848}.

\bibitem[{\citenamefont{Borasoy
  et~al.}(2007{\natexlab{b}})\citenamefont{Borasoy, Epelbaum, Krebs, Lee, and
  Mei{\ss}ner}}]{Borasoy:2007vy}
\bibinfo{author}{\bibfnamefont{B.}~\bibnamefont{Borasoy}},
  \bibinfo{author}{\bibfnamefont{E.}~\bibnamefont{Epelbaum}},
  \bibinfo{author}{\bibfnamefont{H.}~\bibnamefont{Krebs}},
  \bibinfo{author}{\bibfnamefont{D.}~\bibnamefont{Lee}}, \bibnamefont{and}
  \bibinfo{author}{\bibfnamefont{U.-G.} \bibnamefont{Mei{\ss}ner}},
  \bibinfo{journal}{Eur. Phys. J. A} \textbf{\bibinfo{volume}{34}},
  \bibinfo{pages}{185} (\bibinfo{year}{2007}{\natexlab{b}}),
  \eprint{0708.1780}.

\bibitem[{\citenamefont{Lu et~al.}(2016)\citenamefont{Lu, L\"ahde, Lee, and
  Mei\ss{}ner}}]{Lu:2015riz}
\bibinfo{author}{\bibfnamefont{B.-N.} \bibnamefont{Lu}},
  \bibinfo{author}{\bibfnamefont{T.~A.} \bibnamefont{L\"ahde}},
  \bibinfo{author}{\bibfnamefont{D.}~\bibnamefont{Lee}}, \bibnamefont{and}
  \bibinfo{author}{\bibfnamefont{U.-G.} \bibnamefont{Mei\ss{}ner}},
  \bibinfo{journal}{Phys. Lett. B} \textbf{\bibinfo{volume}{760}},
  \bibinfo{pages}{309} (\bibinfo{year}{2016}), \eprint{1506.05652}.

\bibitem[{\citenamefont{Bour et~al.}(2011)\citenamefont{Bour, Li, Lee,
  Mei{\ss}ner, and Mitas}}]{Bour:2011xt}
\bibinfo{author}{\bibfnamefont{S.}~\bibnamefont{Bour}},
  \bibinfo{author}{\bibfnamefont{X.}~\bibnamefont{Li}},
  \bibinfo{author}{\bibfnamefont{D.}~\bibnamefont{Lee}},
  \bibinfo{author}{\bibfnamefont{U.-G.} \bibnamefont{Mei{\ss}ner}},
  \bibnamefont{and} \bibinfo{author}{\bibfnamefont{L.}~\bibnamefont{Mitas}},
  \bibinfo{journal}{Phys. Rev. A} \textbf{\bibinfo{volume}{83}},
  \bibinfo{pages}{063619} (\bibinfo{year}{2011}), \eprint{1104.2102}.

\bibitem[{\citenamefont{Alim and M{\"o}ller}(2009)}]{Alim:2009}
\bibinfo{author}{\bibfnamefont{U.}~\bibnamefont{Alim}} \bibnamefont{and}
  \bibinfo{author}{\bibfnamefont{T.}~\bibnamefont{M{\"o}ller}}, in
  \emph{\bibinfo{booktitle}{International Conference on Sampling Theory and
  Applications (SampTA)}} (\bibinfo{year}{2009}),
  \urlprefix\url{http://eprints.cs.univie.ac.at/4185/}.

\bibitem[{\citenamefont{Zheng and Gu}(2014)}]{Zheng:2014}
\bibinfo{author}{\bibfnamefont{X.}~\bibnamefont{Zheng}} \bibnamefont{and}
  \bibinfo{author}{\bibfnamefont{F.}~\bibnamefont{Gu}},
  \bibinfo{journal}{Journal of Mathematical Imaging and Vision}
  \textbf{\bibinfo{volume}{49}}, \bibinfo{pages}{530} (\bibinfo{year}{2014}).

\end{thebibliography}
\bibliographystyle{apsrev}

\end{document}